\documentstyle[12pt]{article}
\newcommand{\dr}{\mbox{{\footnotesize{$\overline{{\rm DR}}$}}}}
\newcommand{\ms}{\mbox{{\footnotesize{$\overline{{\rm MS}}$}}}}
\newcommand{\gsim}{\buildrel>\over{_\sim}}
\newcommand{\lsim}{\buildrel<\over{_\sim}}
\newcommand{\aBX}{a^*_{Xj}}
\newcommand{\aBY}{a^*_{Yj}}
\newcommand{\bBX}{b^*_{Xj}}
\newcommand{\bBY}{b^*_{Yj}}

\newcommand{\bA}{b_{Xi}}

\newcommand{\aA}{a_{Xi}}

\def\psla{p\kern-.45em/}
\def\dels{\partial\kern-.45em/}
\def\gbar{$\bar{g}_{e\tilde{\nu}\widetilde{W}}$}

\begin{document}

\setcounter{page}{0}
\thispagestyle{empty}
\begin{flushright}
SLAC--PUB--7754 \\
TIT--HEP--383 \\
KEK--TH--562 \\
YITP--98--14 \\
TU--537 \\
\end{flushright}

\vspace{1.7cm}

\begin{center}
\large\bf 
RADIATIVE CORRECTIONS TO A SUPERSYMMETRIC RELATION: A NEW APPROACH\\
\end{center}
\vspace{.8cm}
\baselineskip=32pt
\normalsize
\centerline{Shingo Kiyoura,$^{a,b}$ Mihoko M. Nojiri,$^c$ 
Damien M. Pierce,$^d$ and Youichi Yamada$^e$}

\baselineskip=22pt

\begin{center}
\footnotesize\it
$^a$\,KEK Theory Group, Oho 1-1, Tsukuba, Ibaraki 305-0801, Japan \\

$^b$\,Department of Physics, Tokyo Institute of Technology,
      Oh-okayama, Meguro, Tokyo 152-8551, Japan \\

$^c$\, YITP, Kyoto University, Kyoto 606-8502, Japan \\

$^d$\,Stanford Linear Accelerator Center, Stanford University,
Stanford, California 94309, USA\\

$^e$\,Department of Physics, Tohoku University, Sendai 980-8578, Japan
\end{center}

\vspace{1cm}
\begin{abstract}
Recently it has been realized that the production and decay processes
of charginos, neutralinos, and sleptons receive corrections which grow
like $\log m_{\tilde q}$ for large $m_{\tilde q}$.  In this paper we
calculate the chargino pair production cross section at $e^+e^-$
colliders with quark/squark loop corrections. We introduce a novel
formulation, where the one-loop amplitude is reorganized into two
parts.  One part is expressed in terms of the ``effective'' chargino
coupling \gbar\ and mixing matrices $U^P,\ V^P$, and includes all
${\cal O}(\log m_{\tilde q})$ corrections, while the other decouples
for large $m_{\tilde q}$.  The form of the one-loop cross section then
becomes physically transparent.  Our formulation can be easily
extended to other loops and processes.  Numerically, we find
significant corrections due to the effective $t$-channel coupling
\gbar, for gaugino-like charginos.  In the mixed region, where the
chargino has large gaugino and Higgsino components, the corrections
due to ($U^P$, $V^P$) are also significant.  Our numerical results
disagree with a previous calculation.  We revisit previous studies of
the determination of \gbar\ through the measurement of the chargino
production cross section.  We point out that a previous study, which
claimed that the measurement suffers large systematic errors, was
performed at a ``pessimistic'' point in MSSM parameter space.  We
provide reasons why the systematic errors are not a limiting factor
for generic parameter choices.
\end{abstract}

\vfill


\pagebreak
\normalsize
\baselineskip=15pt
\setcounter{footnote}{0}

\section{Introduction}

Calculations of higher loop effects in the Standard Model (SM),
together with the recent precision measurements of electroweak
parameters, have given rise to a wealth of information on physics at
the weak scale and above.  Among these measurements, one of the
interesting observations is the approximate agreement of the measured
gauge couplings with the prediction of supersymmetric Grand Unified
Theories (GUTs) \cite{GCU}. This may be regarded as indirect evidence
for the minimal supersymmetric standard model (MSSM), which is the low
energy effective theory of supersymmetric GUTs. Additionally, global
fits to precision data in the SM prefer a light Higgs boson mass
\cite{HIGGS}, which is consistent with the MSSM which predicts 
$m_h\lsim 130$ GeV.

In the near future, the Large Hadron Collider (LHC) will explore the
TeV energy region.  Squarks and gluinos will be discovered, together
with charginos and neutralinos, if the supersymmetry breaking scale is
below a few TeV.  Recent studies show that certain superpartner
mass differences can be measured quite precisely at the LHC
\cite{HPSSY,SNOW}. Furthermore, if any of the proposed lepton
colliders are constructed many of MSSM parameters (e.g. the gaugino
masses, the Higgsino mass $\mu$, the slepton masses, and the ratio of
vacuum expectation values $\tan\beta$) will be measured to ${\cal
O}(1\%)\sim {\cal O}(10\%)$ \cite{JLC1,TSUKA,FPMT,NFT}.
 
The future precision measurements of new particle masses, event rates,
and branching ratios, will provide for detailed tests of the
supersymmetry hypothesis. Supersymmetry imposes hard relations between
gaugino couplings and gauge couplings, and between Higgsino and Higgs
couplings.  The cancellation of Higgs mass quadratic divergences
cannot be realized without these supersymmetry coupling relations.
Therefore, they comprise an essential ingredient of the model.
Measurements of the coupling relations will provide definitive
evidence of the realization of supersymmetry in nature.

Because supersymmetry is broken, the hard coupling relations receive
radiative corrections \cite{NFT,CHAN,HN,CFP,RKS,NPY}. 
Since all split supersymmetry multiplets
contribute to the splitting of the gauge/gaugino and Higgs/Higgsino
couplings, measurements of the splitting may provide useful
information about the supersymmetry spectrum. This
is readily understood from the point of view of effective field
theory. As an example, below the squark mass scale the gauge
and gaugino couplings run differently because squarks do not
contribute to the running of the gauge or gaugino couplings, but
quarks continue to contribute to the running of the gauge
couplings. At a scale $Q$ below the squark mass scale, this mismatch
in the running manifests as a difference between the couplings
proportional to $\ln(m_{\tilde q}/Q)$ \cite{CHAN}. Such a correction
also appears in the off-diagonal elements of the chargino and
neutralino mass matrices, which originate from Higgsino-Higgs-gaugino
couplings.  Notice the analogy to the radiative corrections in the
SM. The SM gauge symmetry relates various SM observables. Since the
particle masses in the SM break the gauge symmetry, the measurement of
electroweak observables leads to constraints on the particle masses,
in particular the top quark and Higgs boson masses.

Corrections to supersymmetric relations were first calculated in
Ref.~\cite{CHAN} in the effective renormalization group equation (RGE)
approach. For degenerate squarks, one finds the correction to
lepton-slepton-gaugino couplings
\begin{eqnarray}
\bar{g}_{e\tilde{\nu}\widetilde{W}}/g_2^{\rm SM}
&=& 1+ 2\% \log_{10} (m_{\tilde{q}}/ m_{\tilde{l}})\\
\bar{g}_{e\tilde{e}\widetilde{B}}/g_Y^{\rm SM}
&=& 1+ 0.7\% \log_{10} (m_{\tilde{q}}/ m_{\tilde{l}})~.
\end{eqnarray}

The fermion-sfermion-gaugino couplings are involved in both the
production and decay processes of charginos, neutralinos and
sfermions. Studying these processes provides for measurements of the
gaugino couplings. It is particularly interesting to measure the
gaugino coupling at future $e^+e^-$ colliders, because precise
measurements of the differential cross section are possible there.
Studies show the gauge/gaugino coupling difference may be measured
within $0.3\% \sim 20\%$ through the measurement of the production
cross sections of sleptons or charginos~\cite{NFT,CFP, NPY}.
Typically, the high sensitivity of ${\cal O}(1\%)$ or less may be
achieved when the collider experiments can measure both the final
state superpartner masses and the mass of the particle exchanged in
$t$-channel. Such a high precision measurement of the difference
between the gauge and gaugino couplings allows for the possibility of
constraining the mass scale of squarks which might not be in direct
reach in either hadron or lepton collider experiments.

Because the corrections to the supersymmetric relations are large
enough to be measured in proposed future experiments, it is important
to calculate the full one-loop amplitude in detail.  Tree level
amplitudes depend on the definition of the tree level parameters. In
the \dr\ scheme \cite{DR} the gauge couplings and chargino and 
neutralino mixing matrices depend on the
renormalization scale.  Changing the scale by a factor of 2 easily
results a few percent change in the predicted value of the production
cross section, about the size of the correction of interest. Such
scale dependence can be curtailed only by including radiative
corrections.

In this paper we present the full one-loop calculation of the chargino
production cross section $\sigma(e^-e^+\rightarrow
\tilde{\chi}^-_i\tilde{\chi}^+_j)$ including quark and squark loop
contributions.  The calculation has been performed previously in the
\dr\ scheme in Ref.~\cite{DIAZ}. In their formula the mixing matrix of
the chargino is scale dependent. This scale dependence must be
compensated for by the chargino wave function renormalization, leading
to very complicated expressions. We simplify the calculation by
introducing the effective mixing matrices $U^P, V^P$.  Expressed in
terms of $U^P, V^P$, the formulae are reorganized into a compact and
physically transparent form. This reorganization allows us to see that
the full amplitude consists of two renormalization scale independent
parts.  One contains all the process independent corrections.  For
sufficiently heavy squarks, this reduces to the effective tree level
amplitude which depends on process independent effective parameters.
These effective parameters contain all the corrections proportional to
$\log m_{\tilde{q}}$.  The other part of the amplitude contains the
process dependent contributions, i.e. the one particle irreducible
(1PI) chargino vertex correction and chargino wave function
renormalization. This part decouples in the large $m_{\tilde{q}}$
limit.

In this paper, we also examine previous studies of the measurement of
the effective gaugino coupling \gbar\ through the study of chargino
production and decay \cite{CFP}, which is based on Monte Carlo (MC)
study of Ref.~\cite{FPMT}.  In the study, the constraint on \gbar\ is
claimed to be limited by the systematic error due to the uncertainty
of the underlying parameters. The maximal sensitivity to \gbar\
obtained in Ref.~\cite{FPMT} is 2\%, which is merely enough to
constrain squark mass within a factor of 10. One may ask whether a
full one-loop calculation is necessary if this is always the
case. However, we find the case studied in Ref.~\cite{FPMT} is
uncharacteristically pessimistic in the sense that the signature of
chargino events is very similar to that of backgrounds, and this
naturally makes precision measurement very difficult.  We provide
reasons why systematic errors are not a limiting factor in the
precision study of the supersymmetric relation.

The paper is organized as follows. In Sec.~2 we describe our formalism
which reorganizes the one-loop chargino pair production amplitude,
making for a more physically transparent and computationally
manageable formula. In Sec.~3 we show our numerical results.  The
formulas presented in Sec.~2 systematically guide the discussion of
the $\log m_{\tilde{q}}$ corrections and the remaining finite
corrections.  In addition to corrections to the amplitude from the
well studied $\log m_{\tilde{q}}$ behavior of \gbar, we find the $\log
m_{\tilde{q}}$ corrections and some finite corrections to the
effective mixing matrices are important when the chargino is a sizable
mixture of Higgsino and gaugino.  We also note in Sec.~3 that the
calculation of the cross section including only top-stop and
bottom-sbottom loops presented in Ref.~\cite{DIAZ} is not a reasonable
approximation in general.  Our numerical results disagree with
the results of Ref.~\cite{DIAZ}.  We discuss the validity of various
approximations to the one-loop cross section. In Sec.~4 we revisit the
previous study of the $\bar{g}_{e\tilde{\nu}\widetilde{W}}$
measurement from the chargino production cross section.  In
Ref.~\cite{CFP} they argue that the sensitivity to the cross section
is limited due to a strong dependence of the chargino acceptance on
the theoretical underlying parameters.  However, this study is carried
out at a point in the supersymmetry parameter space with very special
kinematics.  Unlike the generic supersymmetric signature, the signal
has a soft $\psla_T$ distribution similar to the $WW$ background. This
results in the small acceptance and the strong sensitivity to the
masses under the standard set of cuts. We point out that at a generic
point in parameter space the acceptance error is not large and the
systematic error is not a limiting factor in the measurement.  We also
discuss general ways to minimize the acceptance error. Sec.~5 is saved
for discussion and conclusions.

\section{One-loop correction to chargino pair production}

\subsection{Amplitude and cross section}

We show the form of the amplitudes of the chargino pair production
$e^-(p_1)e^+(p_2)\rightarrow
\tilde{\chi}^-_i(p_3)\tilde{\chi}^+_j(p_4)$ including quark and squark
loop corrections. We start with the tree level amplitude. The
$s$-channel amplitude comes from the exchange of gauge bosons
($\gamma$, $Z$).  The $t$-channel amplitude involves the exchange of
the electron sneutrino $\tilde{\nu}\equiv\tilde{\nu}_e$. Their sum
gives
\begin{eqnarray}
i{\cal M}_{ij}&=&i({\cal M}_{ij}^{(s)}+{\cal M}_{ij}^{(t)}) \nonumber\\
&=& +ie^2\frac{1}{s}
[\bar{u}(p_3)\gamma^{\mu}\delta_{ij}v(p_4)]
[\bar{v}(p_2)\gamma_{\mu}u(p_1)] \nonumber\\
&& +ig_Z\frac{1}{s-M_Z^2}
[\bar{u}(p_3)\gamma^{\mu}(v_{Lij}^Z P_L+v_{Rij}^Z P_R)v(p_4)]
[\bar{v}(p_2)\gamma_{\mu}(I_{3e}P_L-s_W^2Q_e)u(p_1)] \nonumber\\ 
&&+\frac{ig_2^2}{2}\frac{V_{i1}^*V_{j1}}{t-m_{\tilde{\nu}}^2}
[\bar{u}(p_3)\gamma^{\mu}P_R v(p_4)]
[\bar{v}(p_2)\gamma_{\mu}P_L u(p_1)]~. \label{eq1}
\end{eqnarray}
where $s=(p_1+p_2)^2$, $t=(p_1-p_3)^2$, $g_Z=g_2/c_W$, $s_W =
\sqrt{1-c_W^2} = \sin\theta_W$, and $P_{R,L}=(1\pm\gamma_5)/2$. The
$u$ and $v$ are the wave functions of $e^{\pm}$ and
$\tilde{\chi}^{\pm}$.  We have applied a Fierz transformation to the
$t$-channel amplitude to make its spinor structure similar to
$s$-channel one. The $v_{ij}^Z$ are the tree level couplings of the
charginos to the $Z$ boson, which depend on the chargino mixing
matrices ($U$, $V$). Their definitions are given in Appendix A.

We next show the form of the corrected amplitude.  The loop
corrections include the 1PI chargino/gauge vertex correction, the
chargino wave function renormalization, and the gauge boson self
energy corrections.  We adopt the \dr\ renormalization scheme for
gauge couplings and the weak mixing angle ($e$, $g_2$, $g_Z$, $s_W$)
and chargino mixing matrices ($U$, $V$), and we adopt the on-shell
scheme for the $Z$ and $\tilde{\nu}$ masses.  Since we only
include the quark and squark loop corrections, the running of the
\dr\ parameters includes only the contributions of quarks and
squarks for consistency.  Note that ($U$, $V$) are obtained by
diagonalizing the tree level mass matrix Eq.~(\ref{eqa2}) in the \dr\
scheme, so they are renormalization scale dependent.

The $t$-channel amplitude receives only chargino wave function
renormalization. The corrected amplitude takes the following form
\begin{eqnarray}
i{\cal M}_{ij}^{(t)}&=&
\frac{ig_2^2}{2}\frac{1}{t-m_{\tilde{\nu}}^2}
[\bar{u}(p_3)\gamma^{\mu}P_R v(p_4)]
[\bar{v}(p_2)\gamma_{\mu}P_L u(p_1)] \nonumber \\
&& \times 
(V_{i1}^*V_{j1}
+\frac{1}{2}(V_{i1}^*V_{j'1}\delta Z^R_{j'j}
+\delta Z^{R\dagger}_{ii'}V_{i'1}^*V_{j1}))~,
\label{eqa9}
\end{eqnarray}
where $\delta Z^{L(R)}$ are the wave function renormalization of
charginos $\tilde{\chi}^-_{L(R)}$. Their explicit forms are given
later.  Strictly speaking, the squark loop correction also appears in
the sneutrino propagator.  However, this momentum independent
contribution is completely canceled by the on-shell renormalization of
$m_{\tilde{\nu}}^2$.

The $s$-channel amplitude is corrected by oblique gauge boson
propagator corrections, chargino wave function renormalization, and
the 1PI chargino-chargino-gauge boson vertex correction. The form of
the corrected amplitude is
\begin{eqnarray}
i{\cal M}_{ij}^{(s)}&=&
-ie Q_e\frac{1}{s}
\left(1- \frac{\Pi^T_{\gamma\gamma}(s)}{s}\right)
[\bar{u}(p_3)\Gamma^{\gamma\mu}_{ij}v(p_4)]
[\bar{v}(p_2)\gamma_{\mu}u(p_1)] \nonumber\\
&&-ie\frac{\Pi^T_{\gamma Z}(s)}{s(s-M_Z^2)}
\biggl\{ Q_{\tilde{\chi}^-}\delta_{ij}
[\bar{u}(p_3)\gamma^{\mu}v(p_4)]
[\bar{v}(p_2)\gamma_{\mu}g_Z(I_{3e}P_L-s_W^2Q_e)u(p_1)] 
\nonumber\\
&&
+Q_e[\bar{u}(p_3)\gamma^{\mu}(v^Z_{Lij}P_L+v^Z_{Rij}P_R)v(p_4)]
[\bar{v}(p_2)\gamma_{\mu}u(p_1)] \biggr\} \nonumber\\
&& -ig_Z\frac{1}{s-M_Z^2}
\left( 1-\frac{\Pi^T_{ZZ}(s)-\Pi^T_{ZZ}(M_Z^2)}{s-M_Z^2} \right) 
\nonumber \\
&&\times 
[\bar{u}(p_3)\Gamma^{Z\mu}_{ij}v(p_4)]
[\bar{v}(p_2)\gamma_{\mu}(I_{3e}P_L-s_W^2Q_e)u(p_1)]~,
\label{eqa10}
\end{eqnarray}
where $M_Z$ is the $Z$-boson pole mass.  The $\Pi^T(p^2)$ are the
transverse parts of the \dr\ renormalized gauge-boson
self-energies. Their explicit forms are given in Ref.~\cite{GB,P}.
The form factors $i\Gamma^{G\mu}_{ij}$ for one-loop corrected
$\tilde{\chi}^+_i\tilde{\chi}^-_jG^{\mu}$ vertices ($G=\gamma,Z$) have
the following forms
\begin{eqnarray}
\Gamma^{G\mu}_{ij}&=&-\gamma^{\mu}(v_{Lij}^G P_L+v_{Rij}^G P_R)
\nonumber \\
&&+ F^G_{VL}\gamma^{\mu} P_L+ F^G_{VR} \gamma^{\mu} P_R 
+ F^G_{SL} (p_3-p_4)^{\mu} P_L + F^G_{SR} (p_3-p_4)^{\mu }P_R \nonumber \\
&&
-\frac{1}{2}(v_{Lij'}^G\delta Z^L_{j'j}
+\delta Z^{L\dagger}_{ii'}v_{Li'j}^G) \gamma^{\mu}P_L
-\frac{1}{2}(v_{Rij'}^G\delta Z^R_{j'j}
+\delta Z^{R\dagger}_{ii'}v_{Ri'j}^G) \gamma^{\mu}P_R~.
\label{eqa11}
\end{eqnarray}
The first line of Eq.~(\ref{eqa11}) contains the tree level couplings
(\ref{eqa6}) with ($e$, $g_Z$, $U$, $V$) in the \dr\ scheme.  $F^G_V$
and $F^G_S$ are the one-particle-irreducible (1PI) corrections to the
vertices.  Their explicit forms are given in the Appendix B. The last
line gives the chargino wave function renormalization.

The wave function corrections $\delta Z^{L,R}$ are determined in terms
of the two-point function $iK_{ij}(p)$ of charginos
$\tilde{\chi}^+_i(-p)\tilde{\chi}^-_j(p)$ in the \dr\ mass basis.
$K_{ij}$ is decomposed as
\begin{equation}
K_{ij}(p)=\Sigma^L_{ij}(p^2)\psla P_L+\Sigma^R_{ij}(p^2)\psla P_R
+\Sigma^D_{ij}(p^2)P_L+\Sigma^{D*}_{ji}(p^2)P_R~, \label{eqa12}
\end{equation}
and $\delta Z$ are then fixed by imposing well-known on-shell
renormalization conditions for fermions \cite{onshell}.  The diagonal
parts of $\delta Z$ are
\begin{eqnarray}
\delta Z^L_{ii}=&& -\Sigma_{ii}^{L}(m_i^2)+
\frac{1}{m_i}\left[\Sigma_{ii}^D(m_i^2)-\Sigma_{ii}^{D*}(m_i^2)\right] 
\nonumber\\
&&-m_i^2\left[\Sigma_{ii}^{L'}(m_i^2)+\Sigma_{ii}^{R'}(m_i^2)\right]
-m_i\left[\Sigma_{ii}^{D'}(m_i^2)+\Sigma_{ii}^{D'*}(m_i^2)\right]
\nonumber\\
\delta Z^R_{ii}=&&-\Sigma_{ii}^{R}(m_i^2)
-m_i^2\left[\Sigma_{ii}^{L'}(m_i^2)+\Sigma_{ii}^{R'}(m_i^2)\right]
-m_i\left[\Sigma_{ii}^{D'}(m_i^2)+\Sigma_{ii}^{D'*}(m_i^2)\right]~.
\label{eqa13}
\end{eqnarray}
Here $\Sigma'(p^2)=\partial\Sigma(p^2)/\partial p^2$.  The
abbreviation $m_i=m_{\tilde{\chi}^-_i}$ is used for convenience.  The
term proportional to ${\rm Im}\Sigma^D_{ii}$ in $\delta Z^L_{ii}$
comes from our convention to use real $\delta Z^R_{ii}$.  In this
paper we treat only cases where $\Sigma_{ii}^D(m_i^2)$ is real (no CP
violation).  The off-diagonal terms ($i\neq j$) are
\begin{eqnarray}
\delta Z^L_{ij}=\frac{2}{m_i^2-m_j^2}\left[
m_j^2\Sigma_{ij}^L(m_j^2)+ m_im_j\Sigma_{ij}^R(m_j^2) 
+m_i\Sigma_{ij}^D(m_j^2)+m_j\Sigma_{ji}^{D*}(m_j^2)
\right]~,
\nonumber\\
\delta Z^R_{ij}=\frac{2}{m_i^2-m_j^2}\left[
m_im_j\Sigma_{ij}^L(m_j^2)+ m_j^2\Sigma_{ij}^R(m_j^2) 
+m_j\Sigma_{ij}^D(m_j^2)+m_i\Sigma_{ji}^{D*}(m_j^2)
\right]~. \label{eqa14}
\end{eqnarray}
In addition, the pole masses of charginos are given by 
\begin{eqnarray}
m_i({\rm pole})&=& m_i-\frac{1}{2}m_i\left[\Sigma_{ii}^L(m_i^2)
+\Sigma_{ii}^R(m_i^2)\right]-\frac{1}{2}\left[\Sigma_{ii}^{D}(m_i^2)
+\Sigma_{ii}^{D*}(m_i^2)\right]~.\label{eqa15}
\end{eqnarray}

In the corrected amplitude Eqs.~(\ref{eqa9}, \ref{eqa10}), the 
renormalization scale dependence of the \dr\ tree level parameters and 
that of the loop functions exactly cancel to ${\cal O}(\alpha)$. However, 
the cancellation is quite complicated as we will see in the next subsection. 

In the numerical calculation we take the pole masses of gauge bosons
($Z$, $W$), and the standard model \ms\ electromagnetic coupling
$\alpha_{\rm SM}(M_Z)$ as inputs. The \dr\ gauge couplings are
obtained from these parameters as discussed in Ref.~\cite{NPY}.  The
chargino sector is fixed by giving pole masses of two charginos and
$\tan\beta(M_Z)$.

Finally, the spin-averaged differential cross section is written in
terms of the amplitude as
\begin{equation}
\frac{d\sigma}{d\cos\theta} = 
\frac{1}{2s}\,\frac{\bar{\beta}_{\tilde{\chi}}}{16\pi}\, \overline{\sum}
\vert {\cal M } \vert ^2~.
\label{eq2}
\end{equation}
Here $\overline{\sum}$ denotes average over the initial electron and
positron helicities and sum over the final chargino helicities.  We
use the helicity amplitude method \cite{HEL} in the numerical
calculation of the cross section. The relevant formulae are given in
Appendix C.  The phase space factor $\bar{\beta}_{\tilde{\chi}}$ is
given by
\begin{equation}
\bar{\beta}_{\tilde{\chi}}=\frac{1}{s}\sqrt{s^2
-2(m_{\tilde{\chi}^-_i}^2+m_{\tilde{\chi}^+_j}^2) s + 
(m_{\tilde{\chi}^-_i}^2-m_{\tilde{\chi}^+_j}^2)^2}~.  \label{eq3}
\end{equation}
Since a highly polarized electron beam will be available at future
$e^+e^-$ linear colliders, in this paper we often present the cross
section for an initial electron in a helicity eigenstate.  Note that
the chargino masses in $\bar{\beta}_{\tilde{\chi}}$ and in the wave
functions ($u(p_3)$, $v(p_4)$) are the pole masses.

\subsection{On-shell Renormalization of charginos}

The wave function renormalizations of charginos $\delta Z^{L(R)}_{ij}$
appear in the corrected amplitude Eqs.~(\ref{eqa9}, \ref{eqa10}). They
contain ultraviolet divergences from ($\Sigma^L$, $\Sigma^R$,
$\Sigma^D$) and, after \dr\ renormalization, depend on the
renormalization scale $Q$.  This $Q$ dependence cancels the implicit
$Q$ dependence of the \dr\ mixing matrices ($U$, $V$), the gauge coupling
$g_2$ in Eq.~(\ref{eqa9}), and the explicit dependence of the 1PI vertex
corrections in Eq.~(\ref{eqa11}).  However, this cancellation is quite
complicated. For example, in Eq.~(\ref{eqa11}) the $Q$ dependence of the
off-diagonal parts of $\delta Z^L$ cancels both that of $U$ in $v^G_L$
and that of $F^G_{VL}$.  Moreover, the off-diagonal parts of $\delta
Z^{L(R)}$ in Eq.~(\ref{eqa14}) superficially diverge when two chargino
masses become degenerate. Therefore the forms of $\delta Z^{L(R)}$ in
Eqs.~(\ref{eqa9}, \ref{eqa10}) can be inconvenient in real calculations.

In this subsection, we reorganize the contribution of $\delta
Z^{L(R)}$ into a very convenient form, by utilizing the
$Q$-independent effective chargino mixing matrices ($U^P$, $V^P$). The
loop contributions which compensate the running of ($U$, $V$) are then
completely split from other corrections.

We first notice that both the diagonal and off-diagonal parts of the
chargino wave function renormalization $\delta Z^{L(R)}$ can be
implemented by making the following replacements in the couplings of
the tree level amplitude,
\begin{eqnarray}
U^\dagger_{ik} &\rightarrow &
U^\dagger_{ik}+\frac{1}{2}U^\dagger_{ij}\delta Z^L_{jk}~, \nonumber\\
V^T_{ik}  &\rightarrow & V^T_{ik}+\frac{1}{2}V^T_{ij}\delta Z^R_{jk}~.
\label{eq4}
\end{eqnarray}
The corrections Eq.~(\ref{eq4}) are universal in any processes involving
on-shell charginos. Remember that in Eq.~(\ref{eq4}) the mixing matrices
($U$, $V$) diagonalize the \dr\ tree level mass matrix Eq.~(\ref{eqa2}).

The factors in Eq.~(\ref{eq4}) come from the relations between the \dr~
fields in the gauge eigenbasis $\psi_i^-$, the \dr\ fields in the
tree level mass eigenbasis $\tilde{\chi}_i^-$, and the on-shell
renormalized fields $\tilde{\chi}^{-P}_i$,
\begin{eqnarray}
&& \psi_{iL}^- = U^\dagger_{ij}\tilde{\chi}^-_{jL}
=U^\dagger_{ij}(Z^L)^{1/2}_{jk}\tilde{\chi}^{-P}_{kL}~, \nonumber\\
&& \psi_{jR}^- = V^T_{ij}\tilde{\chi}^-_{jR} 
= V^T_{ij}(Z^R)^{1/2}_{jk}\tilde{\chi}^{-P}_{kR}~.  \label{eq5}
\end{eqnarray}

We then introduce the effective mixing matrices of charginos, which
are renormalization scale independent, and rewrite Eq.~(\ref{eq4}) by
these matrices.  We first define the effective mass matrix
$\overline{M}_C(p^2)$ in the \dr\ gauge basis as
\begin{equation}
\overline{M}_C(p^2)= M_C -\tilde{\Sigma}^D(p^2)
-\frac{1}{2}M_C\tilde{\Sigma}^L(p^2)-\frac{1}{2}\tilde{\Sigma}^R(p^2)M_C~.
\label{eq6}
\end{equation}
$\tilde{\Sigma}_{ij}$ are chargino two-point functions
$\psi^+_i\psi^-_j$ in the gauge basis. They are related to $\Sigma$ in
the \dr\ mass basis as
\begin{eqnarray}
\Sigma^L &=& U\tilde{\Sigma}^LU^{\dagger}~, \nonumber\\
\Sigma^R &=& V^*\tilde{\Sigma}^RV^T~, \nonumber\\
\Sigma^D &=& V^*\tilde{\Sigma}^DU^{\dagger}~. \label{eq7}
\end{eqnarray}
$\overline{M}_C$ is diagonalized by two unitary matrices, 
the effective mixing matrices $\overline{U}(p^2)$ and 
$\overline{V}(p^2)$, as 
\begin{equation}
\overline{M}_D(p^2)= \overline{V}^*(p^2) \overline{M}_C(p^2) 
\overline{U}^{\dagger}(p^2)~, \label{eq8}
\end{equation}
where $\overline{M}_D(p^2)={\rm diag}(\overline{m}_i(p^2))$ is a
real diagonal matrix. Note that ($\overline{M}_C$, $\overline{U}$, 
$\overline{V}$, $\overline{M}_D$) are independent of 
the \dr\ renormalization scale $Q$. 

We then give the forms of $(\overline{U},\overline{V})$ and 
$\overline{M}_D$ in terms of two-point functions $\Sigma(p)$ of 
charginos. 
$(\overline{U},\overline{V})$ are expanded as 
\begin{eqnarray}
&&\overline{U}(p^2)=U+\delta U(p^2)=(1+\delta u(p^2))U~, \nonumber \\
&&\overline{V}(p^2)=V+\delta V(p^2)=(1+\delta v(p^2))V~.  \label{eq9}
\end{eqnarray}
Here $\delta u=\delta U\cdot U^\dagger$ and $\delta v=\delta V\cdot
V^\dagger$ must be anti-hermitian from the unitarity of
$(\overline{U},\overline{V})$.  Their diagonal elements must be then
pure imaginary.

The ${\cal O}(\alpha)$ expansion of Eq.~(\ref{eq8}) gives 
\begin{eqnarray}
\overline{M}_D(p^2) &=& (1+\delta v^*(p^2))V^* 
\overline{M}_C(p^2) U^{\dagger}(1+\delta u(p^2)) \nonumber \\ 
&=& M_D -\Sigma^D(p^2)
-\frac{1}{2}M_D\Sigma^L(p^2)-\frac{1}{2}\Sigma^R(p^2)M_D \nonumber \\
&& + \delta v^*(p^2)M_D + M_D\delta u(p^2)~.
\label{eq10}
\end{eqnarray}
The real parts of the diagonal elements of Eq.~(\ref{eq10}) give
$\overline{m}_i=m_i({\rm pole})$ at $p^2=m_i^2$.  The off-diagonal elements
of Eq.~(\ref{eq10}) give the following relations
\begin{eqnarray}
\delta u_{ij}(p^2)&=& -\frac{1}{m_i^2 -m_j^2} 
\left( \frac{1}{2}(m_i^2+m_j^2)\Sigma^L_{ij}
+m_i m_j \Sigma^R_{ij}
+m_i\Sigma^D_{ij}+m_j\Sigma^{D*}_{ji} \right) (p^2)~, \nonumber\\
\delta v^*_{ij}(p^2)&=& -\frac{1}{m_i^2 -m_j^2} 
\left( \frac{1}{2}(m_i^2+m_j^2)\Sigma^R_{ij}
+m_i m_j \Sigma^L_{ij}
+m_j\Sigma^D_{ij}+m_i\Sigma^{D*}_{ji} \right) (p^2)~,
\label{eq11}
\end{eqnarray}
for $i\neq j$. 
Finally, the imaginary parts of diagonal elements give the relation 
\begin{equation}
(-\delta u+\delta v^*)_{ii}(p^2)
=\frac{1}{2m_i}[\Sigma^D-\Sigma^{D*}]_{ii}(p^2)~.   \label{eq12} 
\end{equation}
By convention we set $\delta v_{ii}(p^2)=0$. 
The relations Eq.~(\ref{eq5}) are then rewritten in terms of 
\begin{equation}
U^P_{ij}\equiv\overline{U}_{ij}(m_i^2)~,\;\;\;
V^P_{ij}\equiv\overline{V}_{ij}(m_i^2)~,  \label{eq13} 
\end{equation}
as 
\begin{eqnarray}
\psi_{iL}^- &=& (U^P)^\dagger_{ij}
(\delta_{jk}-\frac{1}{2}\Sigma_{jk}^L(m_k^2))N_k^{1/2}
\tilde{\chi}^{-P}_{kL}~, \nonumber\\
\psi_{iR}^- &=& (V^P)^T_{ij}
(\delta_{jk}-\frac{1}{2}\Sigma_{jk}^R(m_k^2))N_k^{1/2}
\tilde{\chi}^{-P}_{kR}~, \nonumber\\ 
N_k^{1/2} &=& 1 
-\frac{m_k^2}{2}\left[\Sigma_{kk}^{L'}+\Sigma_{kk}^{R'}\right](m_k^2)
-\frac{m_k}{2}\left[\Sigma_{kk}^{D'}+\Sigma_{kk}^{D'*}\right](m_k^2)~.
\label{eq14}
\end{eqnarray}
$N_k^{1/2}$ is the real diagonal finite factor.  The chargino wave
function renormalization is then included by replacing
$U^\dagger_{ik}$ and $V^T_{ik}$ in the couplings of the tree level
amplitude by corresponding factors in Eq.~(\ref{eq14}), as
\begin{eqnarray}
U^\dagger_{ik} &\rightarrow& (U^P)^\dagger_{ij}
(\delta_{jk}-\frac{1}{2}\Sigma_{jk}^L(m_k^2))N_k^{1/2}~, \nonumber\\
V^T_{ik} &\rightarrow& (V^P)^T_{ij}
(\delta_{jk}-\frac{1}{2}\Sigma_{jk}^R(m_k^2))N_k^{1/2}~. \label{eq15}
\end{eqnarray}

The use of effective mixing matrices $(U^P,V^P)$ has several nice
features. First, the superficial singularity of $\delta Z_{ij}$ for
degenerate masses is completely absorbed into $(U^P,V^P)$. We can then
see that the original singularity just reflects the arbitrariness in
the diagonalization of a matrix with degenerate eigenvalues.  The
absence of this singularity is similar to the procedure proposed in
Ref.~\cite{KP}.  Second, the renormalization scale dependence in
Eq.~(\ref{eq14}) only appears in $\Sigma^{L,R}$. The $Q$-dependent
parts of the first equation of Eq.~(\ref{eq14}) become, up to ${\cal
O}(\alpha)$,
\begin{eqnarray}
\psi_{iL}^- &=& (U^P)^\dagger_{ij}
(\delta_{jk}-\frac{1}{2}\Sigma_{jk}^L|_{\rm div})\tilde{\chi}^{-P}_{kL} 
\nonumber \\
&=& (\delta_{ij}-\frac{1}{2}\tilde{\Sigma}_{ij}^L|_{\rm div})
(U^P)^\dagger_{jk}\tilde{\chi}^{-P}_{kL}~.
\label{eq16}
\end{eqnarray}
Eq.~(\ref{eq16}) takes the same form as the SU(2)$\times$U(1)
symmetric renormalization of $\psi_i^-$.  This property is very
convenient both in theoretical considerations of the renormalization
and in numerical calculations.  Note, however, that $(U^P,V^P)$ are
non-unitary at ${\cal O}(\alpha)$, unlike $(U,V)$ and
$(\overline{U},\overline{V})(p^2)$\footnote{The effective matrices are
unitary up to corrections of ${\cal
O}((m_{\tilde\chi_2^2}-m_{\tilde\chi_1^2})/m_{\tilde q}^2$).}.

The wave function corrections in our process are expressed as follows.
By applying the rule in Eq.~(\ref{eq15}), the $s$-channel form factors
$i\Gamma^{G\mu}_{ij}$ in Eq.~(\ref{eqa11}) are rewritten as
\begin{eqnarray}
\Gamma^{G\mu}_{ij}&=&-\gamma^{\mu}N^{1/2}_iN^{1/2}_j (\bar{v}_{Lij}^G
P_L+\bar{v}_{Rij}^G P_R) \nonumber \\
&&+\frac{1}{2}(\bar v_{Lij'}^G\Sigma^L_{j'j}(m_j^2)
+\Sigma^{L\dagger}_{ii'}(m_i^2)\bar v_{Li'j}^G) \gamma^{\mu}P_L +\
\frac{1}{2}(\bar v_{Rij'}^G\Sigma^R_{j'j}(m_j^2)
+\Sigma^{R\dagger}_{ii'}(m_i^2)\bar v_{Ri'j}^G) \gamma^{\mu}P_R
\nonumber \\ &&
+\ (\mbox{\rm 1PI vertex corrections})~, \label{eq17}
\end{eqnarray}
where $\bar{v}_{L(R)}^G$ are obtained from $v_{L(R)}^G$ in
Eq.~(\ref{eqa6}) by replacing $(U,V)$ by $(U^P,V^P)$. The
$Q$-dependence of the second line of Eq.~(\ref{eq17}) exactly cancels
that of the third line.  The $Q$-independence of the rewritten form
factor Eq.~(\ref{eq17}) is thus more transparent than the original
form Eq.~(\ref{eqa11}).  Similarly, the last factor in the $t$-channel
amplitude of Eq.~(\ref{eqa9}) is rewritten as
\begin{eqnarray}
&& V_{i1}^*V_{j1}
+\frac{1}{2}(V_{i1}^{P*}V_{j'1}\delta Z^R_{j'j}
+\delta Z^{R\dagger}_{ii'}V_{i'1}^*V_{j1}) \nonumber \\ 
&& = V^{P*}_{i1}V^P_{j1}N^{1/2}_iN^{1/2}_j
-\frac{1}{2}(V_{i1}^{P*}V^P_{j'1}\Sigma^R_{j'j}(m_j^2)
+\Sigma^{R\dagger}_{ii'}(m_i^2)V_{i'1}^{P*}V^P_{j1})~. \label{eq18} 
\end{eqnarray}

In leaving, we comment that the effective matrix method given here can
be applied to any process involving on-shell charginos, since the
corrections of Eq.~(\ref{eq4}) are universal. This method can also be
extended to other particles with flavor mixing, such as the
neutralinos.

\subsection{Large $M_{\widetilde{Q}}$ limit}

We are interested in the limit where the squark mass
$M_{\widetilde{Q}}$ is much larger than the masses of the charginos,
the sneutrino, and the beam energy.  Some corrections to the chargino
production amplitude do not decouple in this limit but increase as
$\log M_{\widetilde{Q}}$.  This reflects the supersymmetry breaking in
the effective field theory below the squark mass scale \cite{CHAN}.

First, the effective chargino mass matrix $\overline{M}_C(p^2)$
receives non-decoupling corrections.  For $M_{\widetilde Q}^2\gg p^2$ the
effective mass matrix becomes independent of $p^2$.  The asymptotic
form is obtained by replacing the elements of the tree level mass
matrix $M_C$ of Eq.~(\ref{eqa2}) by
\begin{eqnarray}
M_2 & \rightarrow & M_2(Q)\left[ 1+ \frac{9g_2^2}{16\pi^2}
\left(\ln\frac{M_{\widetilde{Q}}}{Q}-\frac{1}{4}\right)\right] 
\equiv M_2^{\rm eff}~, \\
\mu & \rightarrow & \mu(Q)\left[ 1+ \frac{3(y_t^2+y_b^2)}{16\pi^2}
\left(\ln\frac{M_{\widetilde{Q}}}{Q}-\frac{1}{4}\right)\right]
\equiv \mu^{\rm eff}~, \\
\sqrt2M_W\cos\beta & \rightarrow & \sqrt2M_W\cos\beta(Q)\biggl[
1 + {\delta M_W\over M_W} + {\delta\cos\beta\over\cos\beta}\biggr]~,
\label{MCbar}\\
\sqrt2M_W\sin\beta & \rightarrow & \sqrt2M_W\sin\beta(Q)\biggl[
1 + {\delta M_W\over M_W} + {\delta\sin\beta\over\sin\beta}\biggr]~,
\end{eqnarray}
where
\begin{eqnarray}
16\pi^2{\delta\cos\beta\over\cos\beta} &=& {3\over2}\sin^2\beta
\left(y_t^2\ln{M_W^2\over Q^2} - y_b^2\ln{M_{\widetilde Q}^2\over
Q^2}\right) - {3\over2}y_b^2\cos^2\beta \ln{M_{\widetilde Q}^2
\over M_W^2} + {9\over4}y_b^2~,\\
16\pi^2{\delta\sin\beta\over\sin\beta} &=& -{3\over2}\cos^2\beta
\left(y_t^2\ln{M_{\widetilde Q}^2\over Q^2} -
y_b^2\ln{M_W^2\over Q^2}\right) -
{3\over2}y_t^2\sin^2\beta\ln{M_{\widetilde Q}^2\over M_W^2} 
+ {9\over4}y_t^2~,\\
16\pi^2{\delta M_W\over M_W} &=& {3\over2}g_2^2\left(
\ln{M_{\widetilde Q}^2\over M_W^2}+{11\over12}\right) \label{dmw}\\
&&-{1\over4}g_2^2\biggl[R(R+2) 
- (3R-2)\ln(R-1) + R^3\ln{R-1\over R}\biggr]~,\nonumber
\end{eqnarray}
with $R=m_t^2/M_W^2$. The corrections to the diagonal elements can be
absorbed into the effective mass parameters $M_2^{\rm eff}$ and
$\mu^{\rm eff}$ and are not interesting within the context of the
MSSM. By contrast, the corrections to the gaugino-Higgsino mixing
masses cannot be absorbed into unknown parameters such as $\tan\beta$.
The squark loop corrections to the effective mass matrix
$\overline{M}_C$, and effective mixing matrices ($U^P$, $V^P$), do not
decouple in the large $M_{\widetilde{Q}}$ limit. This effect is very
important if the gaugino-Higgsino mixing is not highly suppressed.

In the $s$-channel amplitude all other squark loop corrections
decouple in this limit.  The squark loop corrections from the gauge
boson self energies $\Pi^T$ decouple after the gauge couplings are
renormalized.  The factor $N^{1/2}$ in Eq.~(\ref{eq14}) approaches to
1 in this limit. Finally, the non-decoupling terms in $\Sigma^{L(R)}$
in Eq.~(\ref{eq17}) exactly cancel the $F^G_{VL(R)}$ terms of the 1PI
vertex corrections in Eq.~(\ref{eqa11}).  This result is consistent
with the universality of gauge boson interactions.

By contrast, the ${\cal O}(\log M_{\widetilde{Q}})$ terms in
$\Sigma^{L(R)}$ remain in the $t$-channel amplitude and are very
important. This is the origin of the ``super-oblique corrections''
discussed in Refs.~\cite{CFP,RKS,NPY}. We point out that the corrected
$t$-channel amplitude takes a very simple form for sufficiently heavy
squarks.  In this case, the corrected amplitude is obtained from the
tree level one by the replacement
\begin{eqnarray}
g_2^2(Q)V_{i1}^*V_{j1}& \rightarrow & 
g_2^2(Q)
(1-\frac{1}{2}\tilde{\Sigma}^R_{11}(m_i^2)
-\frac{1}{2}\tilde{\Sigma}^R_{11}(m_j^2))V^{P*}_{i1}V^P_{j1}
\nonumber \\
& \equiv & \bar{g}_{e\tilde{\nu}\widetilde{W}}(m_i^2)
\bar{g}_{e\tilde{\nu}\widetilde{W}}(m_j^2)V^{P*}_{i1}V^P_{j1}~.
\label{eq20}
\end{eqnarray}
Here we used the fact that $\tilde{\Sigma}^R_{1j'}(m_j^2)$ ($j'\neq
1$) is insignificant for sufficiently heavy squarks.  The parameter
$\bar{g}_{e\tilde{\nu}\widetilde{W}}$, which is renormalization scale
independent, is interpreted as the effective
$e\tilde{\nu}\widetilde{W}$ coupling.
$\bar{g}_{e\tilde{\nu}\widetilde{W}}$ deviates from the corresponding
gauge coupling, $g_2^{\rm SM}(Q)$. Its asymptotic form is
\cite{CFP,NPY},
\begin{equation}
\bar{g}_{e\tilde{\nu}\widetilde{W}} = g_2^{\rm SM}(Q)\left[
1 + {3g_2^2\over32\pi^2}\left(\ln{M_{\tilde Q}^2\over Q^2}-{3\over4}
\right)\right]~. \label{eq21}
\end{equation}
Here we note that the $m_i^2$ dependence of 
$\bar{g}_{e\tilde{\nu}\widetilde{W}}(m_i^2)$ decouples 
in the large $M_{\widetilde Q}$ limit. 

\section{Numerical results}

In this section we describe the dependence of the chargino production
cross section $\sigma(e^-e^+ \rightarrow
\tilde{\chi}^-_1\tilde{\chi}^+_1)$ on various MSSM parameters.  The
production cross section is a function of the gauge couplings, the
$Z$-boson and sneutrino masses, and the chargino masses and mixing
matrices (parameterized by $M_2$, $\mu$, $M_W$ and $\tan\beta$).  In
the proposed colliders the electron beam can be highly polarized,
therefore we often show the production cross section with a polarized
electron beam.  We denote the cross section as $\sigma_{L(R)}$ when
the initial state electron is left handed (right handed).

In the gaugino region ($M_2\ll|\mu|$), $\tilde{\chi}^-_1$ is
wino-like, and the amplitude receives both $t$-channel and $s$-channel
contributions, unless the initial electron is right handed. If the
electron is right handed, the $t$-channel amplitude vanishes because
of the absence of a $\widetilde{W}e_R\tilde{\nu}$ coupling. 
In the opposite
limit, $|\mu|\ll M_2$, the lightest chargino is Higgsino-like.  Since
the Higgsino couplings to the first and second generation (s)leptons
are negligible, in the Higgsino limit only the $s$-channel amplitude
contributes.  Finally, when $M_2\sim|\mu|$, both charginos have large
gaugino and Higgsino components, and they are somewhat degenerate in
mass.  In this region of parameter space the chargino mixing matrices
relevant in the production cross section are sensitive functions of
$\tan\beta$, which enters in the off-diagonal elements of the chargino
mass matrix.

Formulas for the one-loop corrected chargino production cross section
are given in the previous section and the Appendix B, including quark
and squark loop effects.  The $t$-channel amplitude depends on the
effective coupling \gbar, the effective chargino mixing matrices $V^P$,
and decoupling corrections. The $s$-channel amplitude depends on the
usual gauge couplings, the effective mixing matrices $U^P, V^P$, and
decoupling corrections such as the 1PI gauge-chargino-chargino vertex
correction.

Both amplitudes depend on squark masses $m_{\tilde{q}_i}$, squark
mixing angles $\theta_{\tilde{q}_i}$, and
quark-squark-gaugino(Higgsino) couplings. In this section, we present
our results assuming a universal soft breaking squark mass
$M_{\widetilde{Q}}$ and a universal trilinear coupling $A$ at the weak
scale.  These parameters, along with $\mu$ and $\tan\beta$, determine
the squark masses and mixing angles.  The third generation
quark-squark-Higgsino couplings depend on the top and bottom Yukawa
couplings $y_t$ and $y_b$.  As shown in Eqs.~(\ref{MCbar}--\ref{dmw}),
the heavy top quark can give rise to a sizable correction proportional
to $y_t^2\ln M_{\widetilde{Q}}$, which enters in the off-diagonal
elements of the effective chargino mass matrix. The Yukawa couplings
are also involved in the 1PI vertex corrections when the final state
chargino is Higgsino-like.  The top Yukawa coupling is very large when
$\tan\beta\rightarrow 1$ while $y_b$ is substantial when $\tan
\beta\gsim30$.

In the following we will consider $\tilde\chi_1^-\tilde\chi_1^+$
production in the three cases where the lightest chargino is
predominantly gaugino, predominantly Higgsino, and in the mixed
region. We will refer to the parameter sets listed in
Table~\ref{psets}. We now discuss the three regions in turn.

\begin{table}[htb]
\begin{center}
\begin{tabular}{|c|l|c|c|c|c|c|c|c|} \hline
$\Biggl.$name & description & $m_{\tilde\chi_1^-}$ & $m_{\tilde\chi_2^-}$ & 
$\tan\beta(M_Z)$ & $m_{\tilde\nu}$ & $A$ & $\sqrt s$ & sgn $\mu$
\\[2mm]\hline
$\Biggl.$G1
   & gaugino region,  $|\mu|>M_2$  & 200 & 800 & 2 & 100 & 0 & 500 &$-1$\\[1mm]
H1 & Higgsino region, $|\mu|<M_2$  & 200 & 800 & 2 & 100 & 0 & 500 &$-1$\\[1mm]
G2 & gaugino region,  $|\mu|>M_2$  & 172 & 512 & 4 & 240 & 0 & 500 &$-1$\\[1mm]
H2 & Higgsino region, $|\mu|<M_2$  & 172 & 512 & 4 & 400 & 0 & 500 &$-1$\\[1mm]
M  & mixed region,    $|\mu|<M_2$  & 172 & 255 & 4 & 240 & 0 & 500 &$-1$\\[1mm]
\hline
\end{tabular}
\end{center}
\caption{\small Five parameter sets. All entries with mass units are in GeV.}
\label{psets}
\end{table}

\subsection{Gaugino region}

\begin{figure}[tb]
\vbox{\kern3.7cm\includegraphics{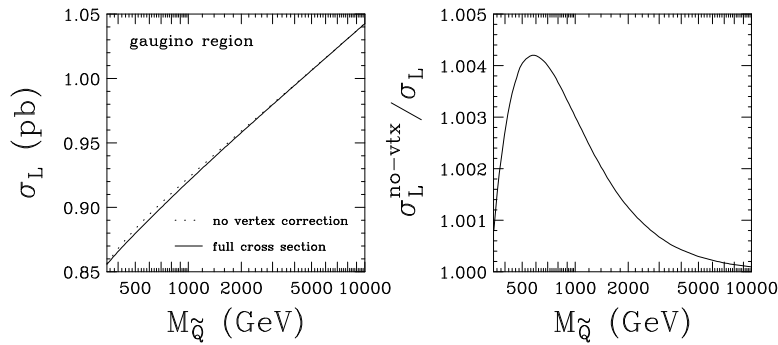}}
\caption{\small (a) The one-loop chargino production cross section
$\sigma_L$ as the function of the soft squark mass $M_{\widetilde Q}$
for the gaugino-like parameter set G1 of Table~1 (solid line). The
positive correction proportional to $\log M_{\widetilde{Q}}$ is due to
the loop correction to \gbar. The dotted line shows the cross section
without the $Z(\gamma)\tilde{\chi}^-\tilde{\chi}^+$ vertex
corrections. (b) The ratio between the cross section without the gauge
vertex corrections and the full one-loop cross section for the same
set of parameters. The vertex correction is less than 0.5\% of the
total cross section.}
\label{fig.cs.g1}
\end{figure}

In Fig.~\ref{fig.cs.g1}(a) we plot the chargino production cross
section $\sigma(e^-_L e^+\rightarrow\tilde{\chi}^-_1
\tilde{\chi}^+_1)$ versus $M_{\widetilde{Q}}$ (solid line), for the G1
gaugino region parameter set of Table~\ref{psets}.  In the gaugino (or
Higgsino) region the diagonal elements of the effective chargino mass
matrix $\overline{M}_C$ are fixed by the input chargino masses, so
they are independent of $M_{\widetilde Q}$. Conversely, the \dr\ 
parameters $M_2(M_2)$ and $\mu(\mu)$ vary as $M_{\widetilde{Q}}$
increases.  The effective mixing matrices $U^P,\ V^P$ contain
non-decoupling $\log(M_{\widetilde Q})+{\rm constant}$
corrections. These corrections arise from corrections to the
off-diagonal elements of the effective chargino mass matrix given in
Eqs.~(\ref{MCbar}--\ref{dmw}). They contribute in both the $s$- and
$t$-channel amplitudes.  However, the dependence of the mixing
matrices on the off-diagonal elements of the effective chargino mass
matrix is suppressed for this set of parameters, so $V^P_{11},\
U^P_{11}\simeq1$ over the whole range of squark mass shown.  The
positive correction proportional to $\log M_{\widetilde{Q}}$ in
Fig.~\ref{fig.cs.g1}(a) is therefore primarily due to the loop
correction to the effective coupling \gbar.

The remaining corrections vanish in the large $M_{\widetilde Q}$
limit. These remaining corrections can be divided up into oblique and
non-oblique parts, each of which satisfies decoupling. The non-oblique
part consists of the 1PI vertex correction and the associated chargino
wave function renormalization. In the following when we refer to the
vertex correction we mean this combination. The vertex correction is
somewhat complicated, so it is worthwhile checking whether this gauge
and scale invariant correction can be neglected. The cross section
calculated without including the vertex correction is shown by the
dotted line in Fig.~\ref{fig.cs.g1}(a), and the ratio between the
cross section without the vertex correction and the full one-loop
cross section, $\sigma_L^{\rm no-vtx}/\sigma_L$, is shown in
Fig.~\ref{fig.cs.g1}(b). The maximum effect of the vertex correction
is less than 0.5\% of the total cross section for this choice of
parameters. The vertex correction is negligible compared to the
sensitivity to $\sigma_L$ in future experiments.

\begin{figure}[tb]
\vbox{\kern4.5cm\includegraphics{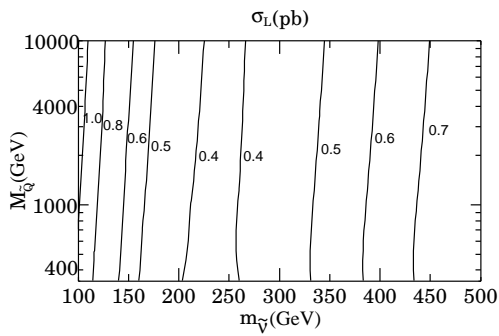}}
\caption{\small Contours of constant $\sigma_L$ in the
$(M_{\widetilde{Q}}, m_{\tilde{\nu}})$ plane for the parameter set G2
of Table 1.  $\sigma_L$ increases (decreases) with $M_{\widetilde Q}$
if $m_{\tilde{\nu}}\lsim 200$ GeV ($\gsim 300$ GeV).}
\label{fig.cs.g2}
\end{figure}

As seen in Fig.~\ref{fig.cs.g1}(a), the left-handed cross section
$\sigma_L$ increases by about 14\% as $M_{\widetilde{Q}}$ varies from
300 GeV to 3 TeV. The sensitivity to $M_{\widetilde Q}$ depends on
$m_{\tilde{\nu}}$ and $\sqrt{s}$, as discussed in Ref.~\cite{CFP}. In
Fig.~\ref{fig.cs.g2} we show contours of constant $\sigma_L$ in the
($m_{\tilde{\nu}}$, $M_{\widetilde{Q}}$) plane, for the G2 parameter
set of Table~\ref{psets}. As $M_{\widetilde Q}$ increases, $\sigma_L$
increases if $m_{\tilde{\nu}}$ is less than 200 GeV, while it
decreases if $m_{\tilde{\nu}}$ is greater than 300 GeV. For
$m_{\tilde{\nu}} \sim 250$ GeV, $\sigma_L$ becomes insensitive to
$M_{\widetilde{Q}}$. The dependence on $M_{\widetilde{Q}}$ from the
$t$-channel amplitude is negligible in the limit
$m_{\tilde{\nu}}\gg\sqrt s$ since the $t$-channel amplitude scales as
$1/m^2_{\tilde\nu}$.

\subsection{Higgsino region}

\begin{figure}[tb]
\vbox{\kern3cm\includegraphics{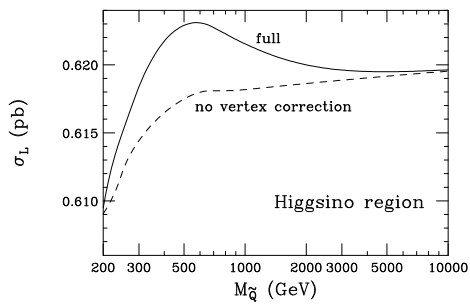}}
\caption{\small The cross section $\sigma_L$ with/without the gauge
vertex correction vs. $M_{\widetilde{Q}}$ for the parameter set H1 of
Table 1 (solid/dotted).  The lightest chargino is Higgsino-like. Both
the dependence on $M_{\widetilde{Q}}$, and the effect of the vertex
correction is very weak.}
\label{fig.cs.H}
\end{figure}

In Fig.~\ref{fig.cs.H} we show the $M_{\widetilde Q}$ dependence
of $\sigma_L$ when the chargino is Higgsino-like. We take parameter
set H1 of Table~\ref{psets}.  The diagonal elements of the effective
chargino mass matrix $\overline{M}_C$ are fixed by fixing the chargino
masses. As in the gaugino region, the mixing is suppressed,
$|V_{12}|$, $|U_{12}|\simeq1$.  The one-loop cross section including
(not including) the vertex correction is shown by the solid (dotted)
line. The cross section changes by less than 0.5\% as $M_{\widetilde
{Q}}$ varies from 300 GeV to 3 TeV. Such a weak $M_{\widetilde Q}$
dependence in Higgsino-like chargino production is expected from our
observations in Sec.~2.  Although the large top quark Yukawa coupling
is involved, the vertex correction remains small, less than 1\%.

\subsection{Mixed region}

\begin{figure}[tb]
\vbox{\kern5.0cm\includegraphics{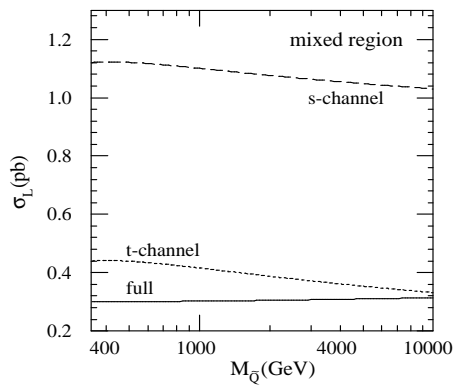}}
\caption{\small $\sigma_L$ vs. $M_{\widetilde{Q}}$ for the mixed
region parameter set M of Table~1 (solid). The $s$-channel and the
$t$-channel cross sections are shown by the long dashed and short
dashed lines. The cross section receives important corrections from
the $\log M_{\widetilde{Q}}$ dependence of $U^P$ and $V^P$ (see
text).}
\label{fig.cs.mix}
\end{figure}

In the mixed region ($M_2\sim|\mu|$), the full one-loop cross section
receives important corrections proportional to $\log M_{\widetilde Q}$
through the corrections to the effective mixing matrices $U^{P}$,
$V^{P}$, as well as $\log M_{\widetilde Q}$ corrections from the
effective coupling \gbar. We illustrate this in Fig.~\ref{fig.cs.mix},
which shows the production cross section for parameter set M of
Table~\ref{psets}. For this choice of parameters $M_2$ and $|\mu|$ are
both near 200 GeV, so the chargino mass eigenstates are fully mixed
($|V^{P}_{11}|^2\simeq0.6$).  In the figure, $\sigma_L$ increases by
4\% as $M_{\widetilde Q}$ varies from 1 to 10 TeV (solid line). The
destructive interference between the $t$-channel and $s$-channel
amplitudes accounts for this insensitivity. The 22\% reduction in the
$t$-channel cross section (short-dashed line) is due to an 8\%
reduction of $V^P_{11}$ and a 2\% increase of \gbar.  The $s$-channel
cross section depends on both $U^{P}$ and $V^{P}$, and decreases by
7\% (long-dashed line).

\subsection{Comparison of $M_{\widetilde Q}$ and $\tan\beta$ dependencies}

\begin{figure}[tb]
\vbox{\kern6.4cm\includegraphics{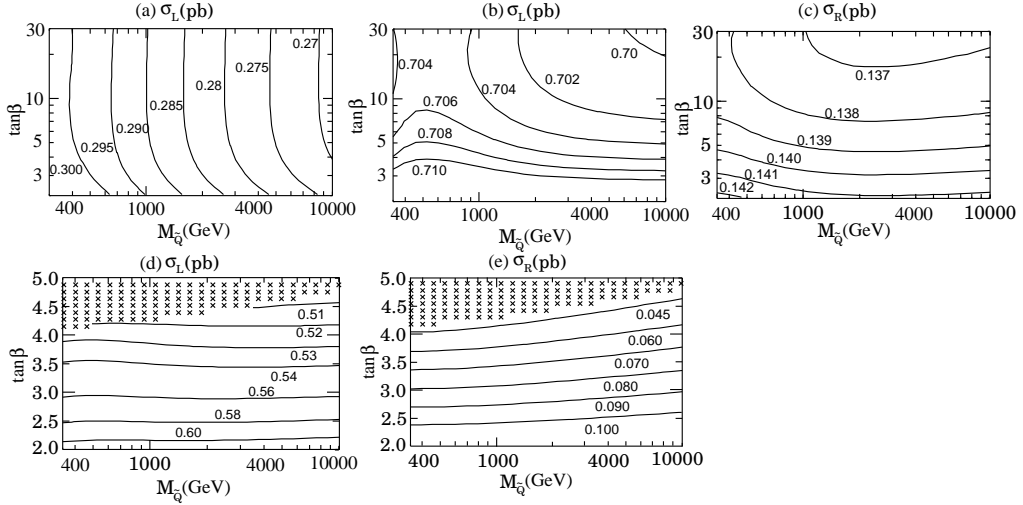}}
\caption{\small Contours of $\sigma_L$ (Figs.~(a), (b) and (d)) and
$\sigma_R$ (Figs.~(c) and (e)) in the $(M_{\widetilde Q},\ \tan\beta)$
plane.  The parameter sets from Table~1 are (a) G2 (gaugino-like) with
$\sqrt{s}=400$ GeV; (b) and (c) H2 (Higgsino-like); (d) and (e) M
(mixed) with $m_{\tilde{\nu}}=400$ GeV. The $M_{\widetilde{Q}}$
dependence is strong in Fig.~(e), though the $\tan\beta$ dependence is
more prominent.  The cross hatched regions are excluded by the
chargino mass constraints.}
\label{fig.cs.con}
\end{figure}

We now compare the $M_{\widetilde Q}$ and $\tan\beta$ dependence of
the chargino production cross section. In
Figs.~\ref{fig.cs.con}(a)--(e), we show contours of constant cross
section in the $(M_{\widetilde{Q}}$, $\tan\beta)$ plane. In
Fig.~\ref{fig.cs.con}(a) we use parameter set G2 of Table~\ref{psets},
with $\sqrt{s}=400$ GeV\footnote{The parameters are those used in
Ref.~\cite{CFP}. We take $\sqrt{s}=400$ GeV for Fig.~\ref{fig.cs.con}(a)
because of the accidental insensitivity of $\sigma_L$ to
$M_{\widetilde{Q}}$ at $\sqrt{s}=500$ GeV.}.  Since $\tilde{\chi}^-_1$
is gaugino-like, $\sigma_L$ depends on $m_{\tilde{\nu}}$. We see that
$\sigma_L$ is insensitive to $\tan\beta$ in this case.  It is almost
constant in $\tan\beta$ when $\tan\beta>5$.  On the other hand,
$\sigma_L$ decreases by 10\% when $M_{\widetilde{Q}}$ changes from 300
GeV to 10 TeV due to the correction to \gbar.

In Figs.~\ref{fig.cs.con}(b) and (c), we show the results for
Higgsino-like $\tilde\chi_1^-$, using parameter set H2 of
Table~\ref{psets}. Over the entire variation of $\tan\beta$ and
$M_{\widetilde Q}$ considered, $\sigma_L$ varies by less than 1.5\%
(Fig.~\ref{fig.cs.con}(b)). The cross section $\sigma_R$ is relatively
more sensitive to $\tan\beta$ (Fig.~\ref{fig.cs.con}(c)), but the
absolute change in the cross section is smaller than in the $\sigma_L$
plot.  In both cases the $M_{\widetilde{Q}}$ dependence is very weak
because of the very small mixing, and the absence of a $t$-channel
coupling.

In the case of large gaugino-Higgsino mixing, the cross section is
more sensitive to $\tan\beta$ than to $M_{\widetilde Q}$. In
Figs.~\ref{fig.cs.con}(d) and (e) we show the chargino production
cross section in the mixed region, with parameter set M of
Table~\ref{psets}, except $m_{\tilde\nu}=400$ GeV. The cross hatched
region $\tan\beta\gsim4$ shown in these plots is excluded because it is
not possible to obtain the specified chargino masses in this
region. We find a strong dependence on $\tan\beta$ for both $\sigma_L$
(Fig.~\ref{fig.cs.con}(d)) and $\sigma_R$ (Fig.~\ref{fig.cs.con}(e)).
The $M_{\widetilde{Q}}$ dependence is very small in Fig.~(d) due to
the interference between the $s$- and $t$-channel amplitudes. In
general it can be large. For example, in Fig.~(e) at $\tan\beta=4$,
$\sigma_R$ changes from 45 fb to 62 fb as $M_{\widetilde{Q}}$ changes
from 300 GeV to 10 TeV.

The large dependence of the one-loop cross section on
$M_{\widetilde{Q}}$ in the mixed region is caused by the strong
sensitivity of the cross section to the off-diagonal elements of the
effective chargino mass matrix.  In Ref.~\cite{FPMT}, it was claimed
that a 4\% measurement of $\sigma_R$ results in the constraint
$3.9<\tan\beta<4.1$ (at $\tan\beta=4$).  The tree level chargino
masses were fixed in their determination. We see from
Fig.~\ref{fig.cs.con}(e) that for a given value of $\sigma_R$ the
central value of $\tan\beta(M_Z)$ can shift by 0.5, depending on
$M_{\widetilde Q}$.

We find that while the vertex correction is not substantial compared
to the experimental sensitivity to the cross section, the corrections
due to $U^P$ and $V^P$ can be.  For example, the $M_{\widetilde Q}$
dependence in Fig.~\ref{fig.cs.con}(e) is almost entirely due to $U^P$
and $V^P$.  At $\tan\beta=4$, $\sigma_R$ changes by about 37\% as
$M_{\widetilde Q}$ varies from 300 GeV to 10 TeV. In contrast, the
vertex correction is less than 1.4\% over this range.

\subsection{Squark mixing effects}

Left-right squark mixing effects give rise to important corrections in
the mixed region. The stop and sbottom mixing angles are controlled by
$A$, $\mu$, and $\tan\beta$ as described in Appendix A. In
Fig.~\ref{fig.cs.sqmix} we show contours of constant $\sigma_R$ in the
($A/M_{\widetilde{Q}}$, $M_{\widetilde{Q}}$) plane for the chargino in
the mixed region.  We use parameter set M of Table~\ref{psets}.  The
cross hatched region of Fig.~\ref{fig.cs.sqmix} is excluded either
because of the chargino mass constraint or because
$m^2_{\tilde{t}}<0$.  The cross section shows strong dependence on
$A/M_{\widetilde{Q}}$ when $M_{\widetilde{Q}}$ is small. For example,
when $A/M_{\widetilde{Q}}$ varies from $-2$ to $+2$ with
$M_{\widetilde{Q}}=345$ GeV, the cross section changes from 61 fb to
39 fb.  The stop mass eigenstates become fully mixed at large
$|A|/M_{\widetilde{Q}}$.  The top squarks are well split when
$A/M_{\widetilde{Q}}=-2$ ($m_{\tilde{t}_1}=128$ GeV,
$m_{\tilde{t}_2}=528$ GeV), while at a point of small mixing,
$A/M_{\widetilde{Q}}=0.12$, the top squarks are nearly degenerate
($m_{\tilde{t}_1}=383$ GeV, $m_{\tilde{t}_2}=386$ GeV). The $A$
dependence of $\sigma_R$ decouples at large $M_{\widetilde{Q}}$ as
$1/M_{\widetilde Q}$ for fixed $A/M_{\widetilde Q}$.

\begin{figure}[tb]
\vbox{\kern3.5cm\includegraphics{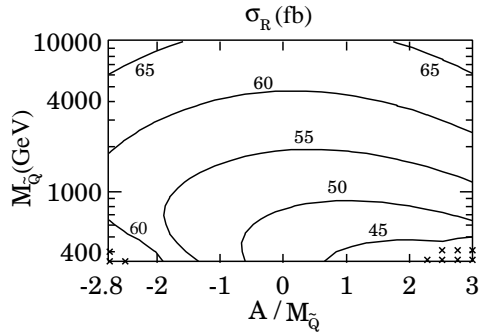}}
\caption{\small Contours of $\sigma_R$ in the $(A/M_{\widetilde Q},
~M_{\widetilde{Q}})$ plane for the mixed region parameter set M of
Table~1. The $A/M_{\widetilde{Q}}$ dependence is larger for smaller
$M_{\widetilde{Q}}$.}
\label{fig.cs.sqmix}
\end{figure}

If only the two chargino masses and $\sigma_R$ are measured it may not
be possible to disentangle the dependence of the cross section on
squark mixing from the dependence on $\tan\beta$ and
$M_{\widetilde{Q}}$.  For example, at $M_{\widetilde Q}=345$ GeV in
Fig.~\ref{fig.cs.sqmix}, we see the cross section is 61 fb at
$A/M_{\widetilde Q}=-2$. We can find the same cross section with the
same chargino masses at $A=0$ by changing $\tan\beta$ from 4 to
3.6. It may be necessary to measure $A$ from other quantities.  The
stop masses and mixing angle can be constrained if
$m_{\tilde{t}_{1,2}}$ and $\sigma(e^-e^+\rightarrow
\tilde{t}_1\tilde{t}^*_1)$ are measured~\cite{DH}. Combining these
measurements with measurements of $\mu$ and $\tan\beta$ from other
processes, $A_t$ can then be determined.

\begin{figure}[tb]
\vbox{\kern5cm\includegraphics{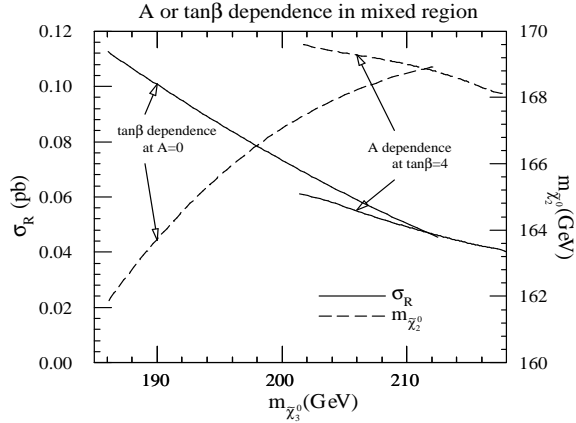}}
\caption{\small Contours of $A=0$ ($\tan\beta=4$) varying $\tan\beta$
($A$) in the $(m_{\tilde{\chi}^0_3},~\sigma_R)$ plane (solid lines),
with $M_{\widetilde{Q}}=345$ GeV. The variation of
$m_{\tilde{\chi}^0_2}$ with $m_{\tilde{\chi}^0_3}$ for the same set of
parameters is also shown (dashed lines).}
\label{fig.A.tanb}
\end{figure}

Because left-right squark mixing arises from $SU(2)\times U(1)$ gauge
symmetry breaking, it contributes to the violations of the relations
between the tree level chargino and neutralino masses. We can utilize
this dependence in efforts to constrain the values of $A$ and
$\tan\beta$. In Fig.~\ref{fig.A.tanb} we show contours of $A=0$
varying $\tan\beta$ and of $\tan\beta=4$ varying $A$ in the
($m_{\tilde{\chi}^0_3}$, $\sigma_R$) plane (solid lines).  We fix
$M_{\widetilde{Q}}=345$ GeV, and the other input parameters are fixed
as in Fig.~\ref{fig.cs.sqmix}.  The contours terminating at
$m_{\tilde{\chi}^0_3}=201.4$ GeV have $\tan\beta=4$ while the contours
terminating at $m_{\tilde{\chi}^0_3}=212.4$ GeV have $A=0$.  For
values of $\tan\beta$ slightly above 4, we cannot find solutions with
the given chargino masses.  For a given $\sigma_R$, we find
$m_{\tilde{\chi}^0_3}$ differs up to 3 GeV between two contours. We
also show the variation of $m_{\tilde{\chi}^0_2}$ with
$m_{\tilde{\chi}^0_3}$ when $A$ or $\tan\beta$ is varied (dashed
lines). For fixed $m_{\tilde{\chi}^0_3}$, $m_{\tilde{\chi}^0_2}$
differs up to 2 GeV between the two curves.  If we can measure the
chargino and neutralino masses within 1 GeV or less, it could help to
single out the effect of squark mixing
\footnote{An excellent measurement of the chargino and neutralino
masses may be achieved at proposed $\mu^+\mu^-$ colliders.  A recent
study shows that it should be possible to measure the lighter chargino
mass with an accuracy of 30 to 300 MeV by measuring the cross section
in the threshold region~\cite{BBH}.}.

Notice the squark mixing dependence of the radiative correction is
mainly due to the correction to $U^P$ and $V^P$, not the vertex
correction.  For example, at $M_{\widetilde Q}=350$ GeV, $\sigma_R$
changes from 47.3~fb to 41.0~fb as $A/M_{\widetilde Q}$ changes from 0
to 2, while the cross section without the vertex correction varies
from 46.7 fb to 40.0 fb, a 1.4\% to 2.4\% effect. Although it is
larger than the $\lsim1\%$ effect found in the gaugino and Higgsino
dominant regions, it is unimportant compared to the strong dependence
of the cross section on $A$, $\tan\beta$, and $M_{\widetilde Q}$.

\subsection{Comparison and approximations}
\setcounter{footnote}{0}

We should briefly comment on the comparison of our results with those
of Ref.~\cite{DIAZ}.  The results in Ref.~\cite{DIAZ} are obtained by
including top, stop, bottom and sbottom loops only.  They
underestimate the $g_2^2\log(M_{\widetilde Q})$ corrections to
\gbar$/g_2$ and $U^P$, $V^P$, which depend equally on the 1st and 2nd
generation (s)quarks as the third generation.  We have checked the
$t$-channel exchange of the sneutrino has a substantial effect on the
total cross section in some of their plots, and including the
contributions of the first two generations significantly alters the
results.

Our comparisons with their results show large numerical differences.
For example, for the parameters corresponding to their Fig.~4 at
$\tan\beta=0.5$, we find the one-loop cross section {\em decreases} by
1.2\% as $M_{\widetilde Q}=A$\footnote{We refer to their $A$. We use the
opposite sign convention for $A$.} varies from 200 to 1000 GeV. Their
results show a 17\% {\em increase} in the cross section.  Notice they
take the gaugino mass parameter $M_2(M_Z)$ as input. Taking this
unphysical mass parameter as input generally leads to larger
$M_{\widetilde Q}$ dependence. In the example just mentioned, taking
$m_{\tilde\chi_2^-}$ as input reduces the $M_{\widetilde Q}$ dependence
from 1.2\% to 0.4\%.

We also find smaller differences between the tree level and one-loop
cross sections. However, this is not surprising, since the definition
of the tree level cross section is somewhat arbitrary.  Our tree level
cross section is determined by the two chargino masses, $M_W$, $M_Z$,
$m_{\tilde\nu}$, $\tan\beta$, and the effective theory (i.e. standard
model) \ms\ gauge couplings. The tree level cross section depends on
the choice of the scale of the effective theory gauge couplings. An
appropriate scale is found by considering the Higgsino production
cross section in the limit $M_Z \ll \sqrt s \ll M_{\widetilde Q}$. In
this limit the squarks are completely decoupled. Because of the
constant correction in the quark loop oblique correction, the cross
section in the effective theory is equal to the cross section in the
full theory if the effective theory gauge couplings are evaluated at
the renormalization scale $Q=\exp(-5/6)\sqrt s\simeq\sqrt s/2$. Hence,
our tree level cross section is evaluated with effective theory gauge
couplings evaluated at the scale $\sqrt s/2$. With this choice, the
tree level and full one-loop cross sections are nearly equal when the
squark corrections decouple.

We have already discussed that, for practical purposes, it is safe to
neglect the vertex correction. We will now consider two approximations
to the remaining corrections. In the first, the effective theory
approximation (ETA), we use the effective coupling \gbar\ and the
effective mixing matrices, $U^P$, $V^P$. In this approximation some
$1/M_{\widetilde Q}^2$ corrections are included. In the second
approximation, the ``log + constant'' approximation (LCA), we strictly
keep only the non-decoupling squark corrections, i.e. we include only
the corrections of the form $\log(M_{\widetilde Q})~+~{\rm constant}$.
The effective coupling in the log + constant approximation, $\bar
g_{\rm LCA}$, is given in Eq.~(\ref{eq21}).  The effective mixing
matrices in the LCA are found as follows. The off-diagonal elements of
the effective chargino mass matrix in the LCA are given in
Eqs.~(\ref{MCbar}--\ref{dmw}).  The effective mixing matrices in the
log + constant approximation, $U^P_{\rm LCA}$, $V^P_{\rm LCA}$, are
then determined from the two chargino masses and these off-diagonal
elements.  Notice that $\bar g_{\rm LCA}$, $U^P_{\rm LCA}$, $V^P_{\rm
LCA}$, and the effective theory couplings, are renormalization scale
independent.

\begin{figure}[tb]
\vbox{\kern4.5cm\includegraphics{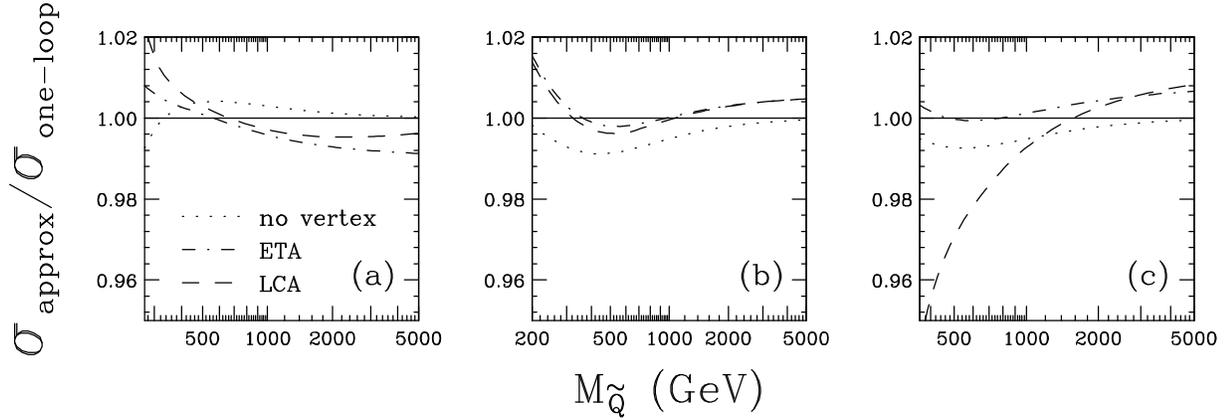}}
\caption{\small Various approximations to the cross section divided by
the full one-loop cross section. The results for the ``no vertex''
approximation, effective theory approximation (ETA), and ``log +
constant'' approximation (LCA) are shown. Figures (a), (b) and (c)
show results in the gaugino, Higgsino, and mixed regions of parameter
space, respectively. (Parameter sets G1, H1, and M of Table~1 are
used.)}
\label{fig.approx}
\end{figure}

In Fig.~\ref{fig.approx} we show the ratio of the various
approximations to the full cross section, versus $M_{\widetilde
Q}$. We plot the ratio of unpolarized cross sections in the gaugino,
Higgsino, and mixed regions in Figs.~\ref{fig.approx}(a), (b), and
(c), with parameter sets G1, H1 and M of Table~\ref{psets},
respectively.  The cross section without the vertex correction is
shown by the dotted line. The ETA result is shown with the dot-dashed
line, and the LCA result is indicated by the dashed line.

There are two factors which contribute to the deviations from unity in
the large $M_{\widetilde Q}$ region in the ETA and LCA results. For
one, these approximations are calculated with effective theory
couplings, while the full calculation is calculated with full theory
couplings. This mismatch causes discrepancies of order $(\alpha\log
M_{\widetilde Q})^2$. These discrepancies give some indication of the
expected magnitude of two-loop corrections. Another reason why the
approximations can disagree in the large $M_{\widetilde Q}$ limit is
that the scale used to evaluate the $s$-channel tree level gauge
couplings is $\sqrt s/2$. The scale which should be used to get exact
agreement in the decoupled regime is somewhat different, depending on
$M_Z$, $m_t$, and $\sqrt s$.

In all three figures the vertex correction is less than 1\%, so the
``no vertex'' approximation is a good one, even for very small squark
masses. The ETA also works well, better than 1\% except at
$M_{\widetilde Q}\lsim\sqrt s/2$ in Fig.~(b).  The LCA works as well
as the ETA, except in the gaugino region with $M_{\widetilde
Q}\lsim\sqrt s$ and in the mixed region with $M_{\widetilde
Q}\lsim1.5\sqrt s$. By comparing the ETA and LCA results we see that
the $1/M_{\widetilde Q}^2$ corrections included in \gbar, $U^P$ and
$V^P$ can be essential in obtaining a good approximation, even for
squark masses as large as $1.5\sqrt s$.

\section{Uncertainty in the chargino production cross section 
measurement}

\subsection{Previous analyses}

In this section we revisit previous studies of chargino production and
decay \cite{FPMT,CFP}. Chargino production can be studied in $e^+e^-$
collider experiments by observing their decay into
$\nu\ell\tilde\chi^0_1$ or $q\bar q'\tilde\chi^0_1$, with signals
$\ell2j$ + missing momentum or $4j$ + missing momentum.

In Refs.~\cite{FPMT,CFP} the probe of the supersymmetric relation of
$SU(2)$ gauge/gaugino couplings $g_2^{\rm SM}=$\gbar\ was considered
based on the MC study of the $\ell2j$ mode at the point in parameter
space $(\mu, M_2, \tan\beta, M_1/M_2, m_{\tilde{\nu}})=(-500$ GeV, 170
GeV, 4, 0.5, 400 GeV), where $m_{\tilde\chi^+_1}=172$ GeV and
$m_{\tilde\chi^0_1}=86$ GeV. The analysis of Ref.~\cite{FPMT} assumes
$\sqrt{s}=500$ GeV and that no direct production of $\tilde{\nu}$ is
available. This results in a poor constraint on \gbar. In
Ref.~\cite{CFP} the authors consider the same point in parameter
space, except they assume $m_{\tilde\nu}$ is measured directly, and
$\sqrt{s}$ can be tuned. Further, they assume that the uncertainty of
the theoretical input parameters in the chargino and neutralino
sector, the acceptance of $\ell2j$ events, and the dependence of the
acceptance on the theoretical input parameters are independent of
$m_{\tilde\nu}$ and $\sqrt s$.  Under these assumptions, they find
$\delta$\gbar$/g_2^{\rm SM}=2\%$ with $m_{\tilde\nu}=240$ GeV and
$\sqrt s=400$ GeV.  This result is considerably poorer than the result
$\delta$\gbar$/g_2^{\rm SM} < 0.6\%$, estimated in the sneutrino
production study of Ref.~\cite{NPY}.

The purpose of this section is to provide a critical discussion of the
analysis of Refs.~\cite{FPMT,CFP}. We point out that the poor
constraint found in Ref.~\cite{CFP} results from the low acceptance of
the $\ell2j$ mode (found in Ref.~\cite{FPMT}), and the low acceptance
is further traced back to the (special) choice of parameters. We also
point out that the signal acceptance and dependence of the acceptance
on the theoretical input parameters can be strongly dependent on
$\sqrt s$. Therefore the estimate given in Ref.~\cite{CFP} cannot be
trusted at other values of $\sqrt s$ unless a dedicated MC simulation
is provided for both the signal and background.  The bottom line is
that the constraint on \gbar\ can be greatly improved for more generic
parameters, and by optimizing the beam energy and cuts.

Before going into the details of their simulation, we shall discuss
the background to the $\ell2j$ signal for the case where $\tilde\chi^{\pm}_1$
decays exclusively into $\tilde\chi^0_1W$ as follows
\begin{eqnarray}
e^+e^-\rightarrow && \tilde\chi^+_1\tilde\chi^-_1      \nonumber\\
                  &&~~~\rightarrow W^+W^-\tilde\chi^0_1
                                         \tilde\chi^0_1\nonumber\\
                  &&~~~~~~~~~~~~\rightarrow l\nu q\bar{q}'
                                    + \ \psla_T~.\nonumber
\end{eqnarray}
We assume $\tilde\chi_1^0$ is the stable LSP, so it escapes detection
and (along with the neutrino) gives rise to missing momentum in these
events.

This process suffers from $W$-boson pair production background. In the
background events the total momentum of the $W$-boson pair is balanced
in the transverse direction, but the observed transverse momentum is
not balanced, due to the escaping neutrino. Hence, the discrimination
between the signal and background is difficult.

In the MC study of Ref.~\cite{FPMT}, the following cuts are made to
reduce the background from $W$-boson pair production in the $\ell2j$
mode:

\vskip12pt
a) existence of an isolated hard lepton; $E_\ell> 5$ GeV,
$\theta_{q\ell}>60\,^\circ$
\vskip12pt

b) $\psla_T> 35$ GeV
\vskip12pt

c) $\theta_{\rm acop}>30\,^\circ$
\vskip12pt

d) $m_{\ell\nu_{\rm ISR}}>120$ GeV
\vskip12pt

e) $-Q_\ell \cos\theta_{\rm had}$, $Q_\ell\cos\theta_\ell<0.707$
\vskip12pt

\noindent Cuts b), c) and d) are set to reduce the $W$-pair events
produced nearly back to back in the transverse direction, while
keeping the supersymmetric signal. The cut e) is designed to remove
the large forward peak of the $WW$ events.

Although these cuts are standard ones to improve the signal to noise
ratio, the acceptance of the signal turns out to be small,
\begin{equation}
\eta={N_{obs}\over\sigma_LB_lB_h{\cal L}}=11.9\%~,
\label{eq:acc}
\end{equation}
resulting in $S/N=1$ at the previously mentioned point in parameter
space.

Our knowledge of the acceptance is limited by the errors of the
underlying parameters $(\mu,$ $M_1,$ $M_2,$ $\tan\beta,$
$m_{\tilde\nu})$. The systematic errors on $\eta$ are estimated as
$\Delta\eta_{\rm sys}=0.55\%$ by allowing the underlying parameters to
vary so that $m_{\tilde\chi^+_1},\ m_{\tilde\chi^+_2},$ and
$m_{\tilde\chi^0_1}$ vary within 2 GeV of their input values.

While $\Delta\eta_{\rm sys}$ itself is small, the error in the cross
section due to the acceptance uncertainty turns out to be large,
i.e. $\Delta\eta/\eta=5\%$. This is comparable to the change in the
cross section when $m_{\tilde q}$ changes from 1 TeV to 4.5 TeV
with $m_{\tilde{\nu}}=240$ GeV and $\sqrt{s}=400$ GeV.

No question was raised concerning the small acceptance of the $\ell2j$
mode in Refs.~\cite{FPMT,CFP}, but the value should be contrasted with
other MC studies, which generally claim acceptances of 30 to 50\% for
SUSY signals \cite{SNOW,TSUKA,BMT}.  The low acceptance in
Eq.~(\ref{eq:acc}) is a consequence of the special choice of
parameters.  At the point in question, $m_{\tilde\chi^+_1}=172$ GeV,
and $m_{\tilde\chi^0_1}=86$ GeV, so the mass difference between the
parent and the daughter particles $\Delta m=m_{\tilde\chi^+_1} -
m_{\tilde\chi^0_1}-m_W$ is only 6 GeV.  In the rest frame of
$\tilde\chi^+_1$, both the $W$-boson and $\tilde\chi^0_1$ are
nonrelativistic. When the parent $\tilde\chi^+_1$ is boosted, the angle
between the $\tilde\chi^+_1$ momentum and the $W$-boson, and also the
momentum spread of the $W$, can be very small.  When the charginos are
produced at $\sqrt{s}= 500 $ GeV, $\theta_{\tilde\chi^+_1 W}<
15^\circ$, and the $W$-boson pair momenta and $\tilde\chi^0_1$
momenta are roughly balanced.

Notice in the cuts listed above we are relying on large
$\theta_{\tilde\chi^+W}$ and large missing transverse momentum to
separate the signal from the $WW$ background, but neither of these
attributes are characteristic of the signal events.  The small
acceptance merely results from the fact that the $W$ pair from the
signal and the background have very similar kinematics at
this particular point in parameter space.

The small acceptance has a direct effect on the acceptance
uncertainty.  The signal event distribution in the $(\psla_T,$
$\theta_{\rm acop},$ $m_{\ell\nu})$ space sits near the background
distribution and therefore near the cut region. When the input
parameters are changed slightly within their error, the signal region
also changes in the $(\psla_T, \theta_{\rm acop}, m_{l\nu})$ space.
Because the accepted number of events for the input parameters is so
small compared to the total number of reconstructed $W$-pair events,
a small change in the signal region easily changes the acceptance by
several percent.  Relatedly, $\theta_{\tilde\chi^+_1 W}$ is a rather
sensitive function of $\Delta m$ when $\Delta m$ is small. Note that
the systematic error of the acceptance is estimated by changing
$\Delta m$ from 2 to 10 GeV in Ref.~\cite{FPMT}.

\begin{figure}[tb]
\vbox{\kern6cm\includegraphics{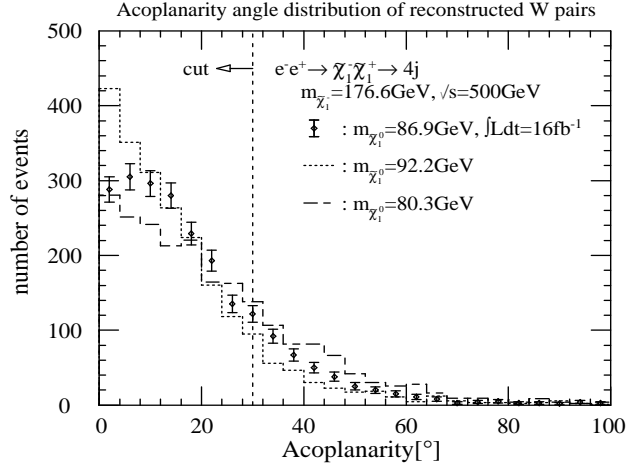}}
\caption{\small The number of accepted chargino production events
vs. the acoplanarity angle, with $\sqrt{s}=500$ GeV and $\int{\cal L}
dt=16$ fb$^{-1}$.  The points with error bars are for $M_1=84.6$ GeV,
and the long-dashed (short-dashed) line corresponds to $M_1=78.6$ (90.6)
GeV. See text for other parameters.}
\label{fig.acop.m}
\end{figure}

To illustrate the kinematics, we show the acoplanarity angle
distribution of $W$ pairs reconstructed from
$e^+e^-\rightarrow\tilde\chi^+_1\tilde\chi^-_1$$\rightarrow
WW\tilde\chi^0_1\tilde\chi^0_1$ $\rightarrow 4j + \psla_T$ events in
Fig.~\ref{fig.acop.m}. The $4j$ mode also suffers from the $WW$
background, however the cuts to remove the SM background are far
simpler than those of the $\ell2j$ mode\footnote{The $4j$ mode suffers
from the SUSY background due to
$e^+e^-\rightarrow\tilde\chi^0_2\tilde\chi^0_2\rightarrow
4j\tilde\chi^0_1\tilde\chi^0_1$, which may be hard to distinguish from
the chargino signal. We discuss the $4j$ mode here for more or less
illustrative purposes, although it is possible to extract more physics
information by including this mode in a combined fit.}.

Fig.~\ref{fig.acop.m} shows the acoplanarity angle distribution
after applying the cuts to reject background $W$-pairs 
given in Ref.~\cite{TSUKA}, except
for the acoplanarity angle cut $\theta_{\rm acop}>30^\circ$.  To
generate MC events, we modified the event generator of
Ref.~\cite{TSUKA}, and we used the JLC detector
simulator~\cite{TSUKA}. The effect of initial state radiation is
included. The distribution shown by points with error bars corresponds
to our standard input parameters $(\mu,$ $M_1,$ $M_2,$ $\tan\beta,$
$m_{\tilde\nu})=(-500$ GeV, 84.6 GeV, 170 GeV, 2, 400 GeV), resulting
in $m_{\tilde\chi^+_1}=176.6$ GeV and $m_{\tilde\chi^0_1}=86.9$
GeV. The parameters are chosen so that $\Delta m=9.4$~GeV, larger than
that in Ref.~\cite{FPMT}. The size of the error bars and the central
values are determined from 10000 generated events, corresponding to
$\int{\cal L} dt=16$ fb$^{-1}$.  On this same plot we also show two
distributions corresponding to $M_1=90.6$ GeV (short-dashed) and
$M_1=78.6$ GeV (long-dashed). These distributions correspond to
$m_{\tilde\chi^0_1}=92.2$ GeV and $m_{\tilde\chi^0_1}=80.3$ GeV,
respectively. The difference in $\Delta m$ for these two curves is 12
GeV, large enough to create a statistically significant difference in
the event distribution, given the somewhat small integrated luminosity
we are considering. These curves are normalized to have equal numbers
of events\footnote{The total number of reconstructed $WW$ events
depends on the chargino/neutralino mass differences. For example,
rejection of the forward going jets ($W$'s) gives such
dependences. However, as discussed below, this mass sensitivity is
small compared to the uncertainty in the acceptance due to the
acoplanarity angle cut.}.

The acceptance changes drastically for the different
cases. Implementing $\theta_{\rm acop}>30^\circ$, we find the
acceptance varies by a factor of two. The acceptance is correlated
with $\Delta m$, which determines the maximal $\tilde\chi^+W$
angle. Below we list $\Delta m$, the maximal angle
$\theta_{\tilde\chi^+W}^{\rm max}$, and the acceptance for the three
cases.
$$\begin{array}{cccc}
\qquad m_{\tilde\chi_1^0} \qquad & \qquad \Delta m \qquad &
\qquad \theta_{\tilde\chi^+W}^{\rm max} \qquad & \qquad \eta \qquad \\[2mm]
\hline\\[-2mm]
92.2 {\ \rm GeV} & 4.1  {\ \rm GeV} & 13.7\,^\circ & 14.1\,\% \\
86.9 {\ \rm GeV} & 9.4  {\ \rm GeV} & 20.7\,^\circ & 19.3\,\% \\
80.3 {\ \rm GeV} & 16.0 {\ \rm GeV} & 27.2\,^\circ & 27.3\,\% \\
\end{array}$$
Varying $m_{\tilde\chi^0_1}$ by 2\%, we expect about a 2\% change in
the acceptance. This corresponds to $\Delta\eta/\eta$ of about 10\%.

Notice we require $\theta_{\rm acop}> 30^\circ$ to reduce the SM
background, while the maximal $\theta_{\rm acop}$ is
$2\theta_{\tilde\chi^+W}^{\rm max}$ if the reconstructed jet momenta
are identified with the quark momenta.  For the sample with $\Delta m=
4.1$ GeV, the events are accepted by virtue of the finite resolution
of the jet axis. As $\theta_{\tilde\chi^+W}$ increases above half of
the acoplanarity angle cut, the accepted number of events increases
linearly with $\theta_{\tilde\chi^+W}$. On the other hand, if $\Delta
m$ is so large that $\theta_{\tilde\chi^+ W}^{\rm max} \gg \theta_{\rm
acop}$, then most of the events pass the cut by a wide margin. In
particular, most events are accepted regardless of several percent
variations in the input parameters. Therefore, the acceptance error is
much smaller in a generic region of parameter space.  We expect the
acceptance uncertainty to scale roughly inversely with the acceptance,
for sufficiently large $\Delta m$. We will examine this conjecture
later by an explicit example. The uncertainty itself depends on the
mode under consideration and the cuts applied, as we discuss in the
next subsection.

For the $\ell2j$ mode the situation is less clear.  Each of the three
cuts, the $\psla_T$ cut, the acoplanarity angle cut, and the
$m_{\ell\nu}$ cut, causes roughly the same reduction in the number of
signal events.  Because of the missing momentum from the escaping
neutrino, each cut yields smaller reductions compared to the $4j$
mode. However, each acceptance dominantly depends on the parameter
$\Delta m$. In particular, the acceptance is larger for larger $\Delta
m$ for all of these cuts.  When $\Delta m$ is small the signal region
significantly overlaps the background region. In that case, as with
the $4j$ mode, the variation of $\Delta m$ within the error is largely
responsible for causing events to move into and out of the accepted
region.

\subsection{Improving the measurement}

We found in the previous subsection that the point in parameter space
considered in Refs.~\cite{FPMT,CFP} must be regarded as a pessimistic
case. At other points the acceptance will generically increase,
leading to a decrease in the acceptance uncertainty. However, nature
may equally well choose any point, so let us reconsider this point for
a moment, and seek a procedure to reduce the acceptance uncertainty.

One possibility is to reduce the chargino and neutralino mass errors,
especially the error on their mass difference.  When there is no
correlation between $\delta m_{\tilde\chi^+_1}$ and $\delta
m_{\tilde\chi^0_1}$, the largest (smallest) acceptance comes from the
point where $m_{\tilde\chi^+_1}-m_{\tilde\chi^0_1}$ becomes maximum
(minimum) within the mass errors. For example, in the $4j$ mode, when
we change $M_1$ and $M_2$ so that $(\Delta m_{\tilde\chi^+_1},\Delta
m_{\tilde\chi^0_1})$ = ($+$2 GeV,$-$2 GeV), we find the acceptance
increases by 6\%. With $(\Delta m_{\tilde\chi^+_1},\Delta
m_{\tilde\chi^0_1})$ =($+$2 GeV,$+$2 GeV) we find the acceptance
increases by 1.7\%\footnote{At ($+$6 GeV, $+$6 GeV) we find the
acceptance is 24.3\%.  We then assume a linear dependence on the
$\Delta m_{\tilde\chi}$ to obtain the estimate for ($+$2 GeV, $+$2
GeV).}. Hence, the acceptance is over three times more sensitive to
absolute changes in the mass difference than the mass sum.

In the standard technique to determine a particle's mass from the
energy distribution of one of the daughter particles in two body
decay, the mass difference between the parent and the daughter
particle is measured better than the individual masses, especially
when the parent particle is significantly boosted. (See examples in
Refs.~\cite{TSUKA,BMT}.) So, at generic points the uncertainty in the
acceptance is much smaller than one would expect from uncorrelated
mass errors.  However, in this particular example, the acceptance is
smaller near the endpoints of the energy distribution. This is because
the daughter particle has a maximal energy when it goes in the same
direction as the parent particle. In such a case, the acoplanarity of
the event comes only from the other chargino, leading to small
statistics near the endpoints.  Therefore, we expect that the energy
distribution is less sensitive to the chargino/neutralino mass
difference for the case given in Ref.~\cite{FPMT}.

\begin{figure}[tb]
\vbox{\kern6cm\includegraphics{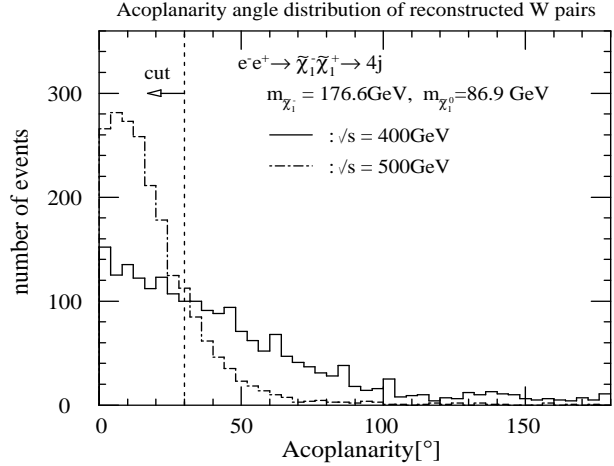}}
\caption{\small The acoplanarity angle distribution of $W$-pair events
from chargino production, at $\sqrt s=400$ GeV. See the text for other
parameters and cuts.}
\label{fig.acop.s}
\end{figure}

The uncertainty of the acceptance may be reduced by increasing the
acceptance itself. One can increase the acceptance easily in the $4j$
mode by reducing $\sqrt s$.  In Fig.~\ref{fig.acop.s} we see the
acoplanarity angle distribution of $W$-pairs is much flatter for
$\sqrt s=400$ GeV. The angle $\theta_{\tilde\chi^+_1W}^{\rm
max}=41.3^\circ$, so a large number of events pass the cut
$\theta_{\rm acop}>30^\circ$. We find the acceptance increases to
54.8\% from 19.3\% at our standard point (where
$m_{\tilde\chi^+_1}=176.6$ GeV and $m_{\tilde\chi^0_1}=86.9$ GeV).
The acceptance increases by 2.4\% with $m_{\tilde\chi^+_1}= 176.6+2$
GeV and $m_{\tilde\chi^0_1}=86.9-2$ GeV. The uncertainty in the
acceptance from the chargino/neutralino mass errors is therefore
reduced by factor of 7 relative to the $\sqrt s=500$ GeV
case\footnote{However, we find the acceptance error including the $WW$
reconstruction efficiency is only reduced by a factor of 2.6. The
dependence of the $W$-pair reconstruction efficiency on the
chargino/neutralino masses comes in through the dependence on the
$W$-boson velocity.  The daughter $W$-boson in the center of mass
frame is substantially nonrelativistic at $\sqrt{s}=400$ GeV, so the
$W$-boson velocity is sensitive to the chargino/neutralino mass
difference. This dependence may be ascertained on an event by event
basis.}.

For the $\ell2j$ mode, reducing $\sqrt{s}$ may not result in a larger
acceptance.  For this mode, many cuts are needed to reduce backgrounds
and the relevant distributions have different $\sqrt{s}$ dependencies.
For example, the $\psla_T>35$ GeV cut rejects more events at smaller
$\sqrt s$, because the narrower allowed range of $\tilde\chi^0_1$
momentum leads to observed events which are more balanced in transverse
momentum. Since the $WW$ background also has a softer $\psla_T$ at smaller
$\sqrt s$, it may be beneficial to reduce the $\psla_T$ cut. To determine
to what extent the acceptance can be improved, both the signal and
background must be studied carefully, because the signal to noise
ratio is near unity in this mode. Such a study is beyond the scope of
this paper.

We summarize this section as follows.

\begin{itemize}
\item The previously claimed cross section error due to the
uncertainty in the acceptance should not be taken as a generic
statement. The point studied in Ref.~\cite{FPMT} has special
kinematic properties making it a very pessimistic case.

\item Because the error in the acceptance scales inversely with the
acceptance, the acceptance uncertainty may be minimized by changing
the beam energy and the cuts so as to maximize the acceptance.  One
should always try to find the best possible way to increase the
acceptance, not in order to increase the statistics, but rather to
reduce the systematic error.

\item If it is clear that specific distributions and/or cuts are the
dominant source of the uncertainty, one might benefit from fitting the
distribution. Fig.~\ref{fig.acop.m} illustrates an example where
the shape of the acoplanarity angle distribution
(not its overall normalization) changes substantially with $\Delta m$.
\end{itemize}

\section{Discussion and conclusions}

In this paper we calculated the chargino production cross section
including full one-loop quark and squark loop corrections.  Quark and
squark loop corrections are known to induce corrections proportional
to $\log m_{\tilde{q}}$.  This logarithmic correction is seen as a
reflection of broken supersymmetry in the effective theory below the
squark mass scale.  The correction may be observable in chargino,
neutralino, and slepton production and decay processes as discussed in
Refs.~\cite{CFP, NPY}. In this paper, the important corrections in the
large $m_{\tilde q}$ limit were extracted from the full one-loop
calculation, and they were compared with the decoupling corrections.
We also revisited previous MC studies of the measurement of the
correction to the fermion-sfermion-chargino coupling $\delta
g=\bar{g}_{e\tilde{\nu}\widetilde{W}}-g^{\rm SM}_2$.  In
Ref.~\cite{CFP} it was stated that a precise determination of the
chargino production cross section is important, but the uncertainty in
the theoretical underlying parameters would be the limiting factor in
the measurement.  In this paper we pointed out the systematic error
will not be a problem at generic points in MSSM parameter space. This
is contrary to their remarks which were based on a MC study at a point
in parameter space with special kinematic properties. Our study shows
that experiments at future $e^+e^-$ colliders should be sensitive to
the squark mass scale if the chargino is produced with a large cross
section.

We presented our one-loop calculation in terms of the renormalization
scale independent effective chargino mixing matrices, $U^P$ and $V^P$.
They are the matrices which diagonalize the effective mass matrix
$\overline{M}_C(p^2)$ at momentum $p^2=m_{\tilde\chi^-}^2$.  When the
one-loop amplitude is written in terms of $U^P$ and $V^P$, a
complicated part of the wave function renormalization is absorbed, and
the remaining part is a simple expression.  The sum of the 1PI
gauge-$\tilde\chi^-\tilde\chi^+$ vertex correction and the remaining
simplified wave function renormalization is scale independent, and
decouples in the large $m_{\tilde q}$ limit.  By isolating this scale
independent correction we were able to discuss its importance
separately.

For sufficiently heavy squarks, it behooves one to introduce the
renormalization scale independent effective electron-sneutrino-wino
coupling \gbar.  All corrections proportional to $\log m_q$ can then
be included in the ``effective tree level'' amplitude, which is
obtained by replacing the couplings and mixing matrices of the tree
level amplitude $g_{e\tilde{\nu}\widetilde{W}}$, $U$, $V$, and
$s$-channel gauge coupling $g_i$, with the effective ones \gbar,
$U^P$, $V^P$, and $g_i^{\rm SM}$ respectively.  The corrections
proportional to $\log m_{\tilde q}$ are included in the first three
effective parameters, while $g^{\rm SM}_i$ is $m_{\tilde{q}}$
independent.

Since we only include quark and squark loop corrections, only the
external chargino lines receive wave function renormalization, and we
discussed the convenience of the introduction of the on-shell
effective chargino mixing matrices in that context.  We note, however,
that our formulation of the effective mixing matrices can be easily
extended to the gauge-Higgs loops, and to the wave function
renormalization of other external particles with flavor mixing, such
as neutralinos.

For gaugino-like charginos, the $\log m_{\tilde q}$ dependence of
\gbar\ gives the dominant correction to the production amplitude.  The
amplitude in Higgsino-like chargino production does not receive
corrections proportional to $\log m_{\tilde q}$. Instead it receives
finite corrections from the gauge-Higgsino-Higgsino vertex correction.
The correction is rather small even though Yukawa couplings are
involved.  Numerically we found the correction is of order a few
percent.  Our numerical calculation is in contradiction with previous
results given in Ref.~\cite{DIAZ}. They claim large corrections to the
production cross section of Higgsino-like charginos. We found in some
cases order of magnitude differences with their results.  Finally, in
mixed chargino production, the corrections to the off-diagonal
elements of the effective chargino mass matrix are important, because
$U^P$ and $V^P$ are sensitive to them.  The off-diagonal elements
receive corrections proportional to $\log m_{\tilde q}$, and they also
receive decoupling corrections due to squark left-right mixing. Both
corrections can be as large as $10\%$.  In order to successfully
extract the most useful information from the chargino measurements, it
may be necessary to isolate the light top squark mixing effects, for
example by measuring the top squark masses and mixing angle through
its direct production. Precise measurements of the chargino and
neutralino spectrum could also give information on top squark mixing.

The validity of various approximations to the one-loop cross section
was also studied in this paper. A simple approximation which works in
a wide region of parameter space makes it easy to simulate the effect
of the radiative correction in MC studies.  We found the 1PI vertex
correction may be safely neglected, and we further defined
approximations to the rest of the one-loop amplitude.  The one-loop
cross section is well described by the effective coupling and mixing
matrices. These parameters encode the leading $\log(M_{\widetilde Q})$
and constant corrections, as well as important decoupling
corrections. If these decoupling corrections are dropped, we found
that the resulting approximation can be poor even for relatively heavy
squark masses, $m_{\tilde{q}} \sim 1.5\sqrt{s}$.

Chargino production suffers from $W$-boson pair production background
at $e^+e^-$ colliders. Therefore, the detectability of the radiative
effect must be studied carefully. Previously studies proceeded by
choosing a point in MSSM parameter space, and generating the MC
signals utilizing the cuts that reduce the $WW$ backgrounds while
keeping signal events. These cuts were determined in a generic
situation in Ref.~\cite{TSUKA}. In Sec.~4, we pointed out that the
point of parameter space chosen in the MC study of Ref.~\cite{CFP} is
not consistent with the assumptions used to determine the cuts in
Ref.~\cite{TSUKA}.  Namely, at the parameter point of Ref.~\cite{CFP}
there is very little phase space in the chargino decay $\tilde\chi_1^-
\rightarrow W\tilde\chi_1^0$. As a result, the $\psla_T$ distribution
of the signal events is similar to that of the background.  However,
the cuts to reduce the background were chosen under the assumption
that the signal would have a higher $\psla_T$ distribution relative to
the background.  We found this causes the very small acceptance.  In
this situation small changes in the $\tilde\chi_1^-$ and
$\tilde\chi_1^0$ masses lead to large variations in the accepted
number of events.  The expected experimental chargino and neutralino
mass error therefore leads to a large systematic error in the chargino
production cross section.  We suggest the uncertainty at such a point
can be reduced by optimizing cuts and beam energy to increase the
acceptance.  We showed that at generic points in
parameter space the acceptance is substantially larger, and the
systematic errors due to the chargino and neutralino mass
uncertainties will not pose a serious problem.  We stress that efforts
to optimize cuts to obtain the maximal acceptance greatly reduce the
error in the cross section both by increasing statistics and reducing
systematic errors.  This improves the sensitivity of the measurement
to the loop effects.

We note the systematic error due to the theoretical underlying
parameters may be reduced by measuring various kinematical
distributions of decay products in $\psla_T$, $\theta_{\rm acop}$,
etc. Such fitting to decay distributions has not been considered in
previous studies.  Furthermore, the decoupling correction is not
negligible in the mixed case, and this might introduce an interesting
twist in future chargino studies. This will be studied elsewhere. We
did not present our fits of chargino production cross section to MSSM
parameters.  Notice, however, that fits of MC data to MSSM parameters
are sensitive to the specific choice of the theoretical input
parameters, beam conditions, etc., chosen for the study.  The fitted
results at one or a few points in parameter space should not be
interpreted generically.

The corrections encoded in \gbar, $U^P$ and $V^P$ are universal.  They
appear in various production and decay processes, and may be important
when chargino decay distributions or branching ratios are used in a
fit.  Neutralino pair production receives analogous $\log{m_{\tilde
q}}$ corrections. Of course the chargino and neutralino corrections
are equally important in final states which receive contributions from
both chargino and neutralino production.

Previously, information on particles which were not produced directly
was ascertained by calculating the effects of loop corrections in SM
processes and comparing the predictions with experimental data.
Unfortunately, superpartners typically give very small corrections in
SM processes because of their decoupling nature. Once a superpartner
is found, the existence of heavier superpartners with mass $M$ gives
rise to interesting non-decoupling effects proportional to $\log M$ in
the production and decay processes of the lighter sparticle.  In this
paper we studied a chargino production process, and compared the $\log
m_{\tilde{q}}$ correction and the associated decoupling corrections in
detail.  We found the mixing of light third generation squarks also
leads important radiative corrections.  If these two effects can be
separated, we could uncover rich information about the squark mass
spectrum. We stress that a systematic treatment of the loop correction
and a detailed examination of future experimental prospects are needed
to make such a study possible.

\section*{Acknowledgements}
We greatly benifitted from discussions with K.~Fujii and appreciate
the support of the JLC group. We thank Ken-ichi Hikasa for bringing
his unpublished note \cite{Hikasa} to our attention. M.M.N. thanks the
SLAC theory group for generous hospitality. S.K. would like to thank
Y.~Okada for helpful discussions.  M.M.N. is supported in part by
Grant in aid for Science and Culture of Japan (07640428, 09246232).
D.M.P. is supported by Department of Energy contract
DE--AC03--76SF00515.

\section*{Appendix A: Tree level interactions}
\setcounter{equation}{0}
\renewcommand{\theequation}{A.\arabic{equation}}

We list the tree level interactions of charginos, quarks, and
squarks. The chargino ($\tilde{\chi}^-$) mass matrix in the gauge
eigenbasis,
\begin{equation}
\psi_{iL}^-=(\widetilde{W}_L^-,\widetilde{H}_{1L}^-)~,\;\;\; 
\psi_{iR}^-=(\widetilde{W}_R^-,\widetilde{H}_{2R}^-)~,  \label{eqa1}
\end{equation}
is given as \cite{GH}
\begin{eqnarray}
-{\cal L}_m= 
\overline{\psi_{iR}^-} M_{Cij} \psi_{jL}^-
+\overline{\psi_{iL}^-} M^\dagger_{Cij} \psi_{jR}^-~,
\nonumber\\
M_C=\left(\begin{array}{cc}
M_2 & \sqrt{2} M_W \cos\beta \\
\sqrt{2} M_W \sin\beta & \mu
\end{array}
\right)~.  \label{eqa2}
\end{eqnarray}
The mass matrix $M_C$ is diagonalized by two unitary matrices $V$ and
$U$ as $M_D= V^* M_C U^{\dagger}$, where $M_D={\rm diag}(m_i)$. Note
that at the one-loop level $M_W$ in Eq.~(\ref{eqa2}) is the \dr\
renormalized parameter.

The chargino-fermion-sfermion couplings are written as follows:
\begin{eqnarray}
{\cal L}_{\rm int}&=&
-\tilde{f_1}_i^*\,\overline{\tilde{\chi}_j^-}
\left(\,a^-_{\tilde{f_1}ij}\,P_L\,+\,b^-_{\tilde{f_1}ij}\,P_R\,\right)\,f_2
+({\rm h.c.}) \nonumber \\
&&-\tilde{f_2}_i^*\,\overline{\tilde{\chi}_j^+}
\,\left(\,a^+_{\tilde{f_2}ij}\,P_L\, + b^+_{\tilde{f_2}ij}\,P_R\,\right)\,f_1
+({\rm h.c.})~, \label{eqa3} 
\end{eqnarray}
where $f=(q,l)$ and $(f_1,f_2)$ are SU(2) doublets, and the suffix $i$
of sfermions denote its mass eigenstates.  Explicit forms of $(a,b)$
in the gauge eigenbasis of sfermions $\tilde{f}_{L,R}$ are written in
terms of $(U,V)$, the gauge coupling $g_2$ and Yukawa couplings $y_f$
of fermions $f$ as
\begin{eqnarray}
&& a^-_{\tilde{f_1}Li}=g_2\,V^*_{i1}~,\; 
a^+_{\tilde{f_2}Li}=g_2\,U^*_{i1}~, \nonumber\\
&& a^-_{\tilde{f_1}Ri}=-y_{f_1}\,V^*_{i2}~,\; 
a^+_{\tilde{f_2}Ri}=-y_{f_2}\,U^*_{i2}~, \nonumber\\
&& b^-_{\tilde{f_1}Li}=-y_{f_2}\,U_{i2}~,\; 
b^+_{\tilde{f_2}Li}=-y_{f_1}\,V_{i2}~,\nonumber \\ 
&& b^{\mp}_{\tilde{f}Ri}=0~,
\label{eqa4}
\end{eqnarray}
where 
\begin{equation}
y_{f_1}=\frac{g_2m_{f_1}}{\sqrt{2}M_W\sin\beta}~,\;\;
y_{f_2}=\frac{g_2m_{f_2}}{\sqrt{2}M_W\cos\beta}~.
\end{equation}
The gauge interactions of fermions, charginos, and sfermions 
are expressed as 
\begin{eqnarray}
{\cal L}_{int}= && -\overline{\tilde{\chi}^-_i}\gamma_{\mu}\, 
\left(v^G_{Lij }P_L +  v^G_{Rij} P_R \right) 
\tilde{\chi}^-_j G^{\mu}\nonumber\\
&&-\bar{f}\gamma_{\mu}\, \left(\, v^G_{fL}\, P_L + 
 v^G_{fR}\, P_R\right) f \,G^{\mu}
\nonumber\\
&&-i\, v^{\tilde{f}G}_{ij}\,(\tilde{f}^*_i \,
\stackrel{\leftrightarrow}{\partial}_{\mu}\, \tilde{f}_j) \, G^{\mu}~. 
\label{eqa5} 
\end{eqnarray}
Here $G=(\gamma, Z)$ and 
\begin{eqnarray}
&& v_{Lij}^G=( U T^G U^{\dagger})_{ij}~, \;\;\;
v_{Rij}^G=( V^* T^G V^T )_{ij}~, \nonumber\\
&&
T^{\gamma}=e\left( \begin{array}{cc} -1 & 0 \\ 
                0 & -1 \end{array} \right)~,\;\;\;
T^Z=g_Z\left( \begin{array}{cc} -1+\sin^2\theta_W & 0 \\ 
                0 & -\frac{1}{2}+\sin^2\theta_W \end{array} \right)~, 
\nonumber\\
&& v^{Z}_{fL}=g_Z\,(\,T_{3fL} -\,Q_f\,\sin^2\theta_W\,)~, 
\;\;\;
v^{Z}_{fR}= - g_Z\, Q_f\,\sin^2\theta_W ~,
\nonumber\\
&& v^{\tilde{f}Z}_{LL}= v^Z_{fL}~, \ \ \ v^{\tilde{f}Z}_{RR}\,= \,
v^Z_{fR}~, \ \ \  v^{\tilde{f}Z}_{LR}\,=\, v^{\tilde{f}Z}_{RL}\,=\, 0~,
\nonumber\\
&& v^{\gamma}_{fL}=v^{\gamma}_{fR}= e\, Q_f~,\ \ \  
v^{\tilde{f}\gamma}_{ij}\,=\,e\,Q_f\,\delta_{ij}~.  \label{eqa6}
\end{eqnarray}

The mixing of left- and right-handed sfermions may not be negligible for 
third generation sfermions. The mass matrices for 
$\tilde{f}=(\tilde{t},\tilde{b})$ are given as follows
\begin{eqnarray}
-{\cal L}_m& =& \left(\begin{array}{cc}\tilde{f}^*_L& \tilde{f}^*_R 
\end{array}\right)\left(
\begin{array}{cc}
m^2_{L}& m^2_{LR} \\
m^2_{LR}& m^2_{R}
\end{array}\right)
\left(\begin{array}{c} \tilde{f}_L \\ \tilde{f}_R\end{array}
\right)~,
\nonumber \\
m^2_L&=&\tilde{m}^2_{\widetilde{Q}_{3L}} + m^2_f + m_Z^2
\cos 2\beta\,(T_{3f_L}\,-\,Q_f\,\sin^2\theta_W)~,
\nonumber\\
m^2_R &=&\tilde{m}^2_{\tilde{f}_{3R}}+ m^2_f + m_Z^2
Q_f\,\cos 2\beta\, \sin^2\theta_W~,
\nonumber\\
m^2_{LR}&=&\left\{\begin{array}{c}
-m_t\,(\,A_t\,+\,\mu\,\cot\beta\,) {\rm \ \ \ \  for} \ \tilde{t}
\\
-m_b\,(\,A_b\,+\,\mu\,\tan\beta\,) {\rm \ \ \ \  for} \ \tilde{b}
\end{array}~.  \right.  \label{eqa7}
\end{eqnarray}
The mass eigenstates $\tilde{f}_{1,2}$ are obtained by 
diagonalizing the mass matrices. This leads to the field rotations
\begin{eqnarray}
\tilde{f}_1&=&\tilde{f}_L\,\cos\theta_{\tilde{f}}
+\tilde{f}_R\,\sin\theta_{\tilde{f}}~,
\nonumber\\
\tilde{f}_2&=&-\tilde{f}_L\,\sin\theta_{\tilde{f}}+
\tilde{f}_R\,\cos\theta_{\tilde{f}}~, \label{eqa8}
\end{eqnarray}
where  $m_{\tilde{f}_1}<m_{\tilde{f}_2}$. $\theta_{\tilde{f}}$ is 
the mixing angle. Couplings of $\tilde{f}_{1,2}$ are easily obtained 
from those in the $\tilde{f}_{L,R}$ basis. 

\section*{Appendix B: Quark and squark loop functions}
\setcounter{equation}{0}
\renewcommand{\theequation}{B.\arabic{equation}}

We list the explicit forms of the quark-squark loop functions in the
corrected amplitude shown in Sec.~2. The results are for quarks and
squarks of a given generation.

The forms of the chargino two-point functions $\Sigma_{ij}(p^2)$ are
\cite{inomasses}
\begin{eqnarray}
\Sigma^L_{ij}(p^2)&=& 
\frac{N_c}{16\pi^2}(
a^{+*}_{\tilde{d}ki}a^+_{\tilde{d}kj}B_1(p^2, m_u, m_{\tilde{d}_k})+
b^-_{\tilde{u}ki}b^{-*}_{\tilde{u}kj}B_1(p^2, m_d, m_{\tilde{u}_k}))~,
\nonumber\\
\Sigma^R_{ij}(p^2)&=& 
\frac{N_c}{16\pi^2}(
b^{+*}_{\tilde{d}ki}b^+_{\tilde{d}kj}B_1(p^2, m_u, m_{\tilde{d}_k})+
a^-_{\tilde{u}ki}a^{-*}_{\tilde{u}kj}B_1(p^2, m_d, m_{\tilde{u}_k}))~,
\nonumber\\
\Sigma^D_{ij}(p^2)&=& 
\frac{N_c}{16\pi^2}(
b^{+*}_{\tilde{d}ki}a^+_{\tilde{d}kj}m_uB_0(p^2, m_u, m_{\tilde{d}_k})+
a^-_{\tilde{u}ki}b^{-*}_{\tilde{u}kj}m_dB_0(p^2, m_d, m_{\tilde{u}_k}))\ . 
\label{eqa16}
\end{eqnarray}
Here $B_{0,1}$ are 't Hooft-Veltman functions in the convention of
Ref.~\cite{P}, and $N_c=3$ is color factor.

The one-particle-irreducible (1PI) corrections to the
$\tilde{\chi}^+_i\tilde{\chi}^-_j G^{\mu}$ vertices, $F^G_V$ and
$F^G_S$, appear in Eq.~(\ref{eqa11})\footnote{There we have ignored
additional terms proportional to $(p_3+p_4)^{\mu}$ since their
contribution vanishes in the massless electron limit. However, these
terms can contribute to other processes such as chargino decays.}.
The corrections have two parts: contributions from
$(f,f,\tilde{f}'_X)$-loops (denoted with $f$) and those from
$(f,\tilde{f}'_X,\tilde{f}'_Y)$-loops (denoted with $\tilde{f}'$),
where $(f,f')$ denotes an SU(2) multiplet of quarks.  Accordingly, the
$F^G$'s are decomposed as
\begin{eqnarray}
F^G_{VL(R)}&=&\sum_f F^{Gf}_{VL(R)}+
\sum_{\tilde{f}}F^{G\tilde{f}}_{VL(R)}~,
\nonumber\\
F^G_{SL(R)} & = & \sum_f  F^{Gf}_{SL(R)}+
\sum_{\tilde{f}}F^{G\tilde{f}}_{SL(R)}~.
\label{eqa17}
\end{eqnarray}

The contribution of the $(f,f,\tilde{f}'_X)$-loops for 
$(f,\tilde{f}')=(d,\tilde{u})$ are expressed as 
\begin{eqnarray}
F^{Gf}_{VL}&=&\frac{N_c}{16\pi^2}\sum_{X=1,2}
\left[ \ \bA \, v_R \, \bBX F^{fX}
+ \aA \,v_L \,\aBX \,m_i\, m_j \,( C^{fX}_{12}
-C^{fX}_{11})\right. \nonumber\\
&&+m_f\left\{ -\aA\, v_R\,\bBX\, m_i\, C^{fX}_{12} - 
\bA \,v_L\, \aBX\, m_j \,C^{fX}_{11} 
\right.\nonumber\\
&&\ \ \ \ \ \ \ \ \left. +\aA\, v_L \,\bBX\, m_i \,
(C^{fX}_0 + C^{fX}_{12}) + 
\bA\, v_R \, \aBX\, m_j \,(C^{fX}_0 + C^{fX}_{11})
\right\}
\nonumber\\
&&\left.+m_f^2 \, \bA \,v_L\, \bBX \,C^{fX}_0\right]~, \label{eqa18}
\\
F^{Gf}_{SL}&=& 
\frac{N_c}{16\pi^2}\sum_{X=1,2}\left[ \bA\, v_R\, \bBX \, m_i\,
( C^{fX}_{22}-C^{fX}_{23} )
\nonumber\right.\\
&&+\aA\, v_L\,\aBX\, m_j\,(C^{fX}_{11}-C^{fX}_{12}
+C^{fX}_{21}-C^{fX}_{23})
\nonumber\\
&&\left. +m_f\,\left\{  \aA\, v_R\,  \bBX\, C^{fX}_{12} - 
 \aA\, v_L \, \bBX\, (C^{fX}_0+ C^{fX}_{11}) \right\} \right]~. \label{eqa20}
\end{eqnarray}
Here we abbreviate $a_{Xi}=a^-_{\tilde{f}'Xi}$,
$b_{Xi}=b^-_{\tilde{f}' Xi}$, and $v_{L(R)}= v^G_{f,L(R)}$. The
$F^{Gf}_R$ formula are obtained from the corresponding $F^{Gf}_L$
expressions by replacing $a_{Xi}\leftrightarrow b_{Xi}$ and
$v_L\leftrightarrow v_R$.  $C^{fX}_0$ and $C^{fX}_{\alpha\beta}$ are
the Passarino and Veltman \cite{PV} $C$ functions in the convention of
\cite{HHKM}. The arguments of the $C$ function are
\begin{equation} 
C^{fX}_{0(\alpha\beta)}= C_{0(\alpha\beta)}(p_j^2, p_i^2, s, m^2_f,
m^2_{\tilde{f}'_X}, m^2_f)~.
\end{equation}
The function $F^{fX}$ is defined as 
\begin{equation}
F^{fX}\equiv -\frac{1}{2}
\left\{ B^{fX} -s C^{fX}_0 + 
(m^2_{\tilde{f}'_X}-m^2_f+ m^2_i-s)\, C^{fX}_{11}
+(m^2_f-m^2_{\tilde{f}'_X}-m^2_i)\, C^{fX}_{12} -1\right\}~,
\end{equation}
where $B^{fX}= B_0(p_j^2, m^2_f, m^2_{\tilde{f}'_X})$.

The contributions of $(f,\tilde{f}'_X,\tilde{f}'_Y)$-loops 
for $(f,\tilde{f}')=(d,\tilde{u})$  are 
expressed as follows 
\begin{eqnarray}
F^{G \tilde{f}'}_{VL}&=&\frac{N_c}{16\pi^2} 
\sum_{XY}2\bA\, v_{XY}\,\bBY \,C^{\tilde{f}'XY}_{24}~, \label{eqa22}
\\
F^{G\tilde{f}'}_{SL} &=& - \frac{N_c}{16\pi^2}\sum_{XY}
\left[m_{f} \,\aA\, v_{XY}\,\bBY \, 
(C^{\tilde{f}'XY}_{12}-C^{\tilde{f}'XY}_{11})
+m_i \,\bA \,v_{XY}\, \bBY\, 
(C^{\tilde{f}'XY}_{23}-C^{\tilde{f}'XY}_{22})\right. 
\nonumber\\ &&\left.
+m_j\, \aA\, v_{XY}\, \aBY \, 
(C^{\tilde{f}'XY}_{23}-C^{\tilde{f}'XY}_{21}+ 
C^{\tilde{f}'XY}_{12}-C^{\tilde{f}'XY}_{11})\right]~.\label{eqa24}
\end{eqnarray}
The $F^{G\tilde f'}_R$ are obtained from
Eqs.~(\ref{eqa22}--\ref{eqa24}) by replacing $a_{Xi}\leftrightarrow
b_{Xi}$.  Here, $v_{XY}= v^{\tilde{f}'G}_{XY}$, and
\begin{equation}
C^{\tilde{f}'XY}_{0(\alpha\beta)}\,=\, 
C_{0(\alpha\beta)}(p_j^2, p_i^2,\, s,\, 
m^2_{\tilde{f}'_Y},\, m_f^2,\,  m^2_{\tilde{f}'_X})~.
\end{equation}
The contributions for $(f,\tilde{f}')=(u,\tilde{d})$ loops can be
obtained from the $(f,\tilde{f}')=(d,\tilde{u})$ expressions in
Eqs.~(\ref{eqa18}--\ref{eqa20}, \ref{eqa22}--\ref{eqa24}) by replacing
$a^-_{\tilde{f}'Xi}\rightarrow b^{+*}_{\tilde{f}'Xi}$,
$b^-_{\tilde{f}'Xi}\rightarrow a^{+*}_{\tilde{f}'Xi}$,
$v^G_{f,L}\leftrightarrow - v^G_{f,R}$, and
$v^{\tilde{f}'G}_{XY}\rightarrow -v^{\tilde{f}'G}_{XY}$.

\section*{Appendix C: Helicity amplitude method}
\setcounter{equation}{0}
\renewcommand{\theequation}{C.\arabic{equation}}

In the calculation of cross sections it is often useful to directly
evaluate the amplitude for helicity eigenstates of initial and final
state particles, and numerically take the sum of the squared
amplitudes for helicities, instead of taking the trace of the squared
amplitude analytically.  This method is called the helicity amplitude
method \cite{HEL,Hikasa}.

In this appendix we list the spinor bilinears which are relevant in
the one-loop amplitude of the process $e^-(p_1, h_1)e^+(p_2,
h_2)\rightarrow\tilde{\chi}^-_i(p_3, h_3)\tilde{\chi}^+_j(p_4, h_4)$.
$h_{1-4}$ are the helicities of the corresponding particles and take
the values $\pm1/2$. We evaluate the spinor bilinears in the center of
mass frame. The coordinate space is chosen so that the initial $e^-$
goes along the positive $z$ axis and the final $\tilde{\chi}_i^-$ goes
along the $(\theta\, \phi)$ direction in the polar basis.  We give our
results in the spherical basis. A Lorentz vector in the spherical
basis, $(A^{\Diamond}, A^m)$ with $m=0,\ \pm$, is related to the
vector $A^\mu$ in the Minkowski basis via
\begin{eqnarray}
A^{\Diamond}&=&A^0~, \nonumber\\
A^+&=& -\frac{1}{\sqrt{2}}\,(A^1+i A^2)~, \nonumber\\
A^0&=&A^3~, \nonumber\\
A^-&=&\frac{1}{\sqrt{2}}\,(A^1-i A^2)~.
\end{eqnarray}
The inner product of two vectors $(A,B)$ is given by
\begin{equation}
A\cdot B \equiv A^{\Diamond} B^{\Diamond}+ \sum_m (-1)^{m+1}A^m B^{-m}~.
\end{equation}

The initial massless fermion bilinears $H$ (we ignore the electron
mass) and final fermion bilinears $\overline H$ are given in the
spherical basis as
\begin{eqnarray}
H_V^{\mu} &\equiv& \bar{v}_{h_2}(p_2)\gamma^{\mu} u_{h_1}(p_1) =
\left(\, 0, 2\sqrt{2}|\lambda_I|E \delta_{-m,\lambda_I}\,\right)~,
\nonumber\\
H_A^{\mu} &\equiv& \bar{v}_{h_2}\gamma^{\mu}\gamma_5 u_{h_1} =
(-1)^{h_2 +1/2}\, H_V^{\mu}~,
\nonumber\\
\overline{H}_S &\equiv&  \bar{u}_{h_3}(p_3) v_{h_4}(p_4) =
\left((E_3+E_4)^2 - (m_i+m_j)^2\,\right)^{1/2}\,\delta_{\lambda_F 0}~,
\nonumber \\ \overline{H}_P &\equiv& \bar{u}_{h_3}\gamma_5 v_{h_4} =
(-1)^{h_4 +1/2}
\left((E_3+E_4)^2 - (m_i-m_j)^2\,\right)^{1/2}\,\delta_{\lambda_F 0}~,
\nonumber \\
\overline{H}^{\mu}_V &\equiv&  \bar{u}_{h_3}\gamma^{\mu} v_{h_4}
= \left( \,
\left[\,\sqrt{E_3+m_i}\,\sqrt{E_4-m_j}\,
-\,\sqrt{E_3-m_i}\,\sqrt{E_4+m_j}\,\right]\,\delta_{\lambda_F 0}, \right.
\nonumber \\ 
& &\ \ \ \ \   \left. - (\sqrt{2})^{|\lambda_F|}
\,\sqrt{\frac{E_3 +m_i}{E_4+m_j}} 
\,\left[\,E_4 +m_j- \,(-1)^{\lambda_F}\,(E_3-m_i)\,\right]
d^1_{m \lambda_F}(\theta) e^{i(m-\lambda_F)\phi}\right)~,
\nonumber\\
\overline{H}^{\mu}_A &\equiv&\bar{u}_{h_3}\gamma^{\mu}\gamma_5v_{h_4}
= (-1)^{h_4 +1/2}\,\left( \,\sqrt{\frac{E_4+m_j}{E_3+m_i}}
\,\left[\,-E_4+E_3 +m_i+m_j\,\right]\,\delta_{\lambda_F 0}, \right. 
\nonumber\\
& & \ \left.-   (\sqrt{2})^{|\lambda_F|} 
\,\left[\, (-1)^{\lambda_F+1}\, 
\sqrt{E_3-m_i}\,\sqrt{E_4+m_j}\, +\,\sqrt{E_3+m_i}\,\sqrt{E_4-m_j}\,\right]
d^1_{m \lambda_F}(\theta) e^{i(m-\lambda_F)\phi}\,\right)~.
\nonumber\\
\end{eqnarray} 
We use the abbreviations $\lambda_I\equiv h_1-h_2$ and
$\lambda_F\equiv h_3-h_4$.  ($E$, $E_3$, $E_4$) are the ($e^-$,
$\tilde{\chi}^-_i$, $\tilde{\chi}^+_j$) energies, respectively.  The
relevant Wigner $d$-functions $d^1_{m\lambda}(\theta)$ are
\begin{equation}
\left( \begin{array}{ccc} 
d^1_{++} & d^1_{+0} & d^1_{+-} \\
d^1_{0+} & d^1_{00} & d^1_{0-} \\
d^1_{-+} & d^1_{-0} & d^1_{--} \end{array} \right) (\theta)= 
\left( \begin{array}{ccc}
\frac{1+\cos\theta}{2} & -\frac{1}{\sqrt{2}}\sin\theta & 
\frac{1-\cos\theta}{2} \\
\frac{1}{\sqrt{2}}\sin\theta & \cos\theta & 
-\frac{1}{\sqrt{2}}\sin\theta \\
\frac{1-\cos\theta}{2} & \frac{1}{\sqrt{2}}\sin\theta & 
\frac{1+\cos\theta}{2} \end{array} \right)~.
\end{equation}

\end{document}